\documentclass[11pt,onecolumn,german]{article}

\usepackage[ngerman]{babel}

\usepackage{mathptmx}
\usepackage[T1]{fontenc}
\usepackage{amsmath}
\usepackage{amssymb}

\newcommand{\changefont}[3]{
\fontfamily{#1} \fontseries{#2} \fontshape{#3} \selectfont}

\usepackage{dsfont}
\usepackage{theorem}
\usepackage{multicol}
\usepackage{graphicx}
\usepackage{epsfig}

\begin{document}
\setcounter{page}{35}
\thispagestyle{empty}
\changefont{ptm}{m}{n}

\noindent {\LARGE{Symmetrien, Teilchen und Felder}}\\
\\

\scshape
\noindent Andreas Aste$^{1,2}$\\

\itshape
\noindent $^{1}$Departement Physik,
Universit\"at Basel,
Klingelbergstrasse 82,
CH-4056 Basel, Schweiz
\\
$^{2}$Paul Scherrer Institut,
CH-5232 Villigen PSI, Schweiz\\

\noindent e-mail: andreas.aste@unibas.ch\\
\\

\normalfont
\noindent{\bfseries{Zusammenfassung:}}
Symmetrien spielen in den Naturwissenschaften eine \"ausserst prominente Rolle. In der Mathematik als
Sprache der Physik werden Symmetrien im Rahmen der Gruppentheorie behandelt, welche zugleich
das n\"otige R\"ustzeug zur Klassifikation von Naturgesetzen wie auch von physikalischen Objekten wie
beispielsweise den Elementarteilchen liefert. In der vorliegenden Arbeit werden die in der
relativistischen Quantenfeldtheorie als mathematische Theorie der Teilchenphysik
verwendeten Begriffe thematisiert, die im Zusammenhang
mit der modernen Beschreibung der bisher als fundamental erachteten Elementarteilchen und der
mit ihnen assoziierten Felder Verwendung finden. Es liegt in der abstrakten Natur der Quantenfeldtheorie,
dass diese Begriffe im engen Rahmen einer \"Ubersichtsarbeit nur durch Veranschaulichung ber\"uhrt werden
k\"onnen.
\\
 
\noindent{\bfseries{Abstract:}}
Symmetries are playing a very prominent r\^ole in natural sciences. In mathematics as the language of
physics, symmetries are treated within the framework of group theory, which provides the tools to
classify natural laws and physical objects like elementary particles. The present work discusses aspects
of relativistic quantum field theory as the mathematical theory of particle
physics which are relevant for the modern description of elementary particles and their
associated fields hitherto considered as fundamental building blocks of the theory.
Due to the abstract nature of quantum field theory, these aspects can only be touched
by their exemplification within a review.
\\


\newpage

\begin{multicols}{2}{
\subsection*{Einf\"uhrung}
Symmetrien (nach den altgriechischen Worten syn, $\sigma \upsilon \nu$ = zusammen und 
metron, $\mu \epsilon \tau \rho \omega \nu$ = Mass)
sind uns aus dem gemeinen Alltag durchaus vertraut. In der entsprechenden Begriffswelt ist die Symmetrie
verkn\"upft mit ebenso intuitiv erfassten Ausdr\"ucken wie Regelm\"assigkeit, Proportion und Harmonie.
Man trifft Symmetrien beim Betrachten von Pflanzen, Kristallen, Kirchenfenstern und Schneeblumen am kalten
Fenster an.

In der Physik kommt dem Symmetriebegriff eine fundamentale, mathematisch klar definierbare
Bedeutung zu. Von der Regelm\"assigkeit
als Qualit\"at eines Subjekts wissenschaftlicher Untersuchungen ist es zur Regel, also dem Naturgesetz,
nicht weit. Entsprechend werden Theorien im Umfeld der Elementarteilchenphysik h\"aufig durch die ihnen
innewohnenden Symmetrien bezeichnet. So gilt beispielsweise die Quantenelektrodynamik, welche die durch elektrische
Ladungen erzeugten Ph\"anomene zum Thema hat, als eine sogenannte abelsche $U(1)$-Eichtheorie,
die Quantenchromodynamik, welche die starken Kr\"afte in Teilchen wie dem Proton oder dem Neutron
beschreibt, ist eine nicht-abelsche $SU(3)$-Eichtheorie. Dabei sind die Terme $U(1)$ und $SU(3)$ Bezeichnungen
f\"ur sogenannte Lie-Gruppen (Lie und Engel 1888), welche in der Mathematik bereits im vorletzten Jahrhundert eingef\"uhrt wurden
und heute zur Klassifikation der Symmetrien von Naturgesetzen und der ihnen unterworfenen Systeme
herangezogen werden.

Es ist ein Ziel dieser Arbeit, Einblicke in die grundlegenden Symmetriekonzepte zu gew\"ahren, die der
Klassifikation der fundamentalen Bausteine der Natur dienen. Da sich die Elementarteilchenphysik und die ihr
zugrunde liegende Theorie, die Quantenfeldtheorie, zwangsl\"aufig abstrakter mathematischer Konzepte bedienen
muss, soll dies mit Hilfe von Veranschaulichungen und Analogiebetrachtungen geschehen. Dem Begriff ``Feld''
kommt in der Quantenfeldtheorie eine recht abstrakte mathematische Bedeutung zu, deren Kl\"arung hier zu weit f\"uhren
w\"urde. Quantenfelder als Grundbausteine der Quantenfeldtheorie sind aber eng mit den mit ihnen assoziierten physikalischen
Ph\"anomenen verkn\"upft, welche in vielen F\"allen einen korpuskularen, also teilchenartigen Charakter aufweisen.

\subsection*{Symmetrien und Gruppen}
Nat\"urlich ist im Rahmen der pr\"azisen mathematischen Formulierung physikalischer Fragestellungen
eindeutig zu kl\"aren, was unter einer Symmetrie \"uberhaupt zu verstehen ist. In der Physik ist es die Eigenschaft
eines n\"aher zu definierenden Systems, unter einer bestimmten Menge von \"Anderungen oder
{\itshape{Transformationen}} invariant zu sein.
Wenn eine Transformation den Zustand oder eine Eigenschaft des Zustandes eines physikalischen oder gedachten
Systems nicht \"andert, werden diese Transformationen Symmetrieoperationen oder eben Symmetrietransformationen
genannt. Unterschieden werden diskrete Symmetrien und kontinuierliche Symmetrien.
\subsubsection*{Diskrete Symmetrien}
Ein Beispiel f\"ur eine diskrete Symmetrie ist die Spiegelungsinvarianz des in Abb.~1 dargestellten Schmetterlings,
einem Tagpfauenauge.
Jeder Farbfleck auf dem linken Fl\"ugel findet eine gespiegelte Entsprechung auf dem rechten Fl\"ugel und umgekehrt.
Man wird nat\"urlich bei genauer Betrachtung einwenden m\"ussen, dass die unterstellte Symmetrie nicht exakt ist.
Diese Situation einer \emph{leicht gebrochenen} oder \emph{approximativen} Symmetrie wird auch in der
Teilchenphysik oft in physikali\-schen Systemen beobachtet und ist nicht unbedingt st\"orend,
sondern ein m\"oglicher Hinweis darauf, dass
ein offensichtlicher symmetrischer Mechanismus durch einen weiteren Mechanismus beeinflusst wird, der selbst wiederum
gewissen Symmetrieprinzipien gehorchen mag. Die pr\"azise Messung solcher kleinen St\"orungen ebnet dann den Weg zum 
Verst\"andnis neuer physikalischer Ph\"anomene und besch\"aftigt gegenw\"artig tausende von Physikern auf der Welt, so zum
Beispiel auf dem sehr aktuellen Gebiet der Neutrinophysik.
\begin{center}
\noindent \includegraphics[width=7.0cm]{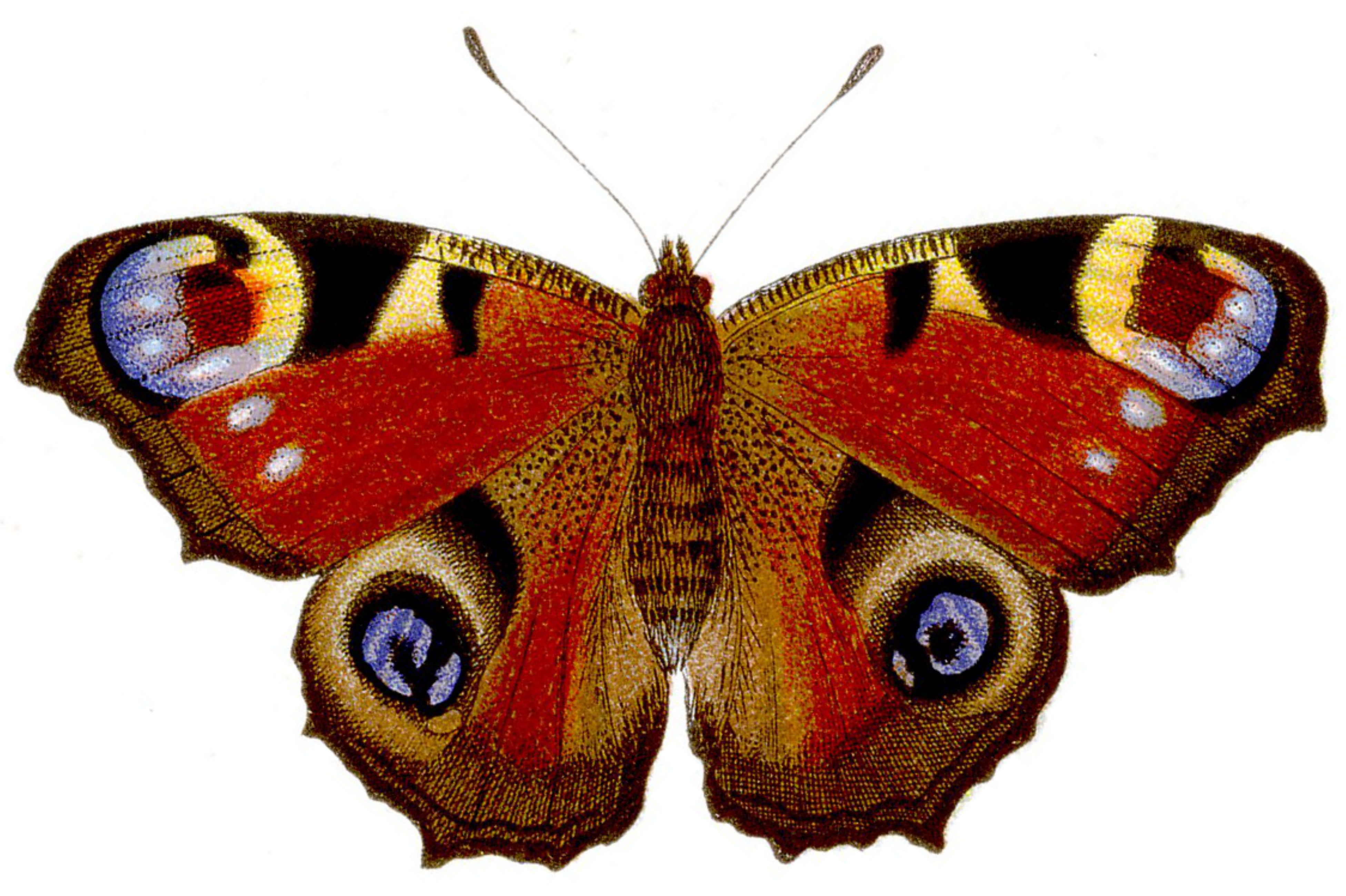}
\end{center}
\noindent {\small{{\bfseries{Abb.~1:}} Tagpfauenauge (Zeichnung: Jacob H\"ubner, um 1800).}}\\

Nach diesem kurzen Exkurs ruhe aber der Brennpunkt unserer Betrachtungen tats\"achlich auf einem idealen, also
perfekt spiegelsymmetrischen Tagpfauenauge, wie es schliesslich durch Abb.~2 dargestellt ist. Das blaue Auge unten links
findet eine genaue Entsprechung im unteren Teil des rechten Fl\"ugels.
Zusammen mit der trivialen Eigenschaft eines Objekts, sich nicht zu ver\"andern, wenn es nicht ver\"andert wird,
existieren also folgende Transformationen, unter welchen das ideale Tagpfauenauge invariant ist:
Die sogenannte Identit\"at $I$, welche nichts bewirkt, und eine Spiegelung $S$ an einer Symmetrieebene.
Es steht uns frei, die Spiegelung zweimal auf das Tagpfauenauge anzuwenden. Durch diese Verkn\"upfung zweier
Symmetrietransformationen erhalten wir offensichtlich die identische Transformation, gar unabh\"angig davon,
ob wir ein reales Tagpfauenauge mit approximativer Spiegelsymmetrie oder ein perfektes Tagpfauenauge zweimal
spiegeln; das Resultat wird dem Urzustand entsprechen. Formal k\"onnen wir die Hintereinanderausf\"uhrung mehrerer
Symmetrietransformationen als Produkt schreiben, sodass also gilt
\begin{equation}
S \cdot S = S^2 = I \, .
\end{equation}
Nat\"urlich gilt weiter
\begin{equation}
I \cdot S = S \cdot I = S \, , \, \, \, I \cdot I =I \, .
\end{equation}
\begin{center}
\noindent \includegraphics[width=7.0cm]{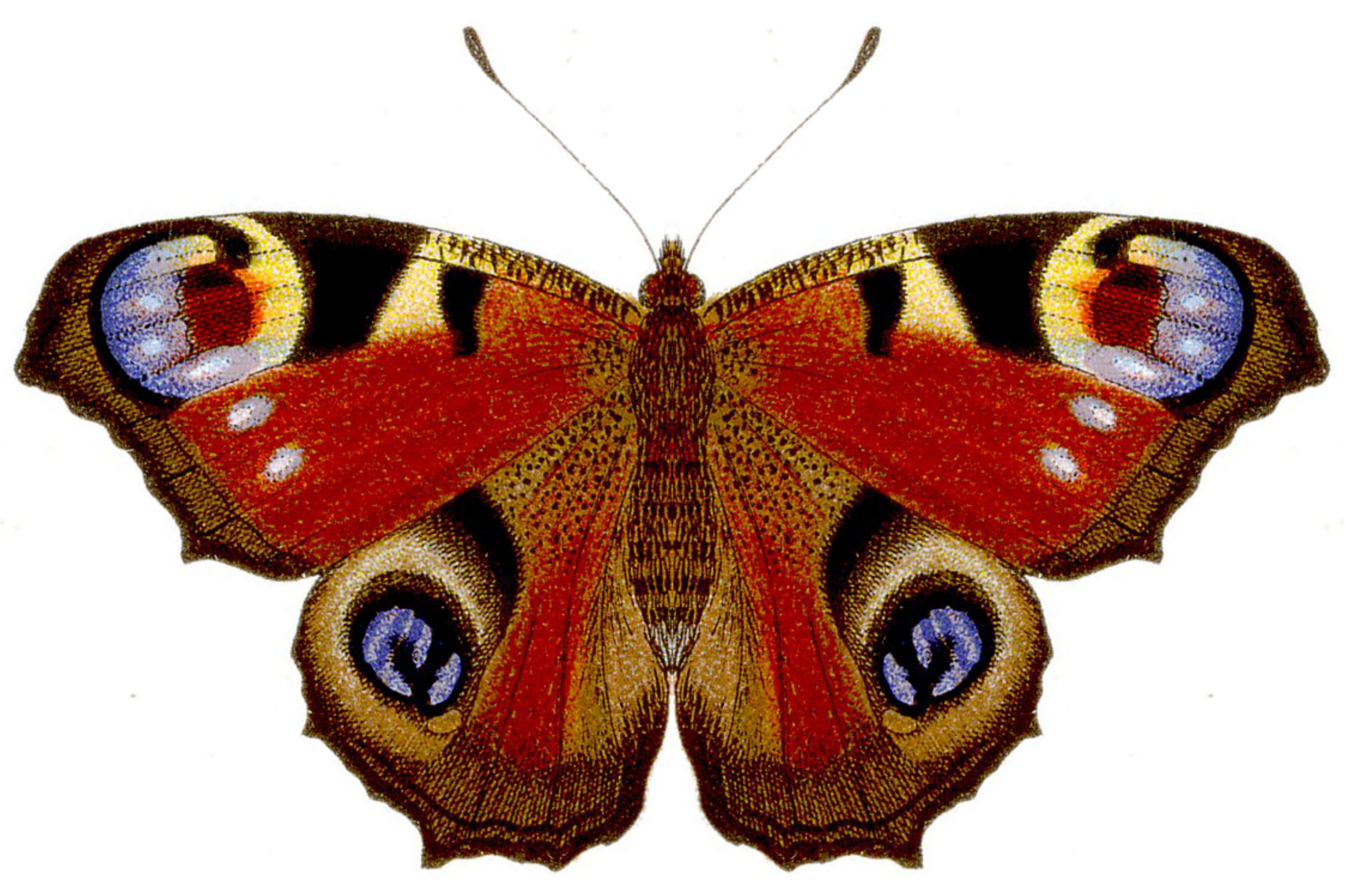}
\end{center}
\noindent {\small{{\bfseries{Abb.~2:}} Exakt spiegelsymmetrisches Tagpfauenauge.}}\\

Damit sind wir beim mathematischen Begriff der \emph{Gruppe} angelangt.
In der Mathematik versteht man unter einer Gruppe eine Menge von Elementen, die paarweise
in geordneter Reihenfolge miteinander verkn\"upft werden k\"onnen und welche
mittels dieser Verkn\"upfung wiederum ein Element
der Gruppenmenge erzeugen. Der vollst\"andige Satz an Forderungen, die an eine Gruppe gestellt
werden, stellt sicher, dass man mit den Gruppenelementen in gewissem Sinne vern\"unftig
rechnen kann. Man definiert also abstrakt:\\

\noindent \emph{Eine Gruppe ist ein Paar $(G,\cdot)$, bestehend aus einer Menge $G$ und einer
Verkn\"upfung ``$\cdot$'' je zweier Elemente aus $G$, also eine Abbildung
\begin{displaymath}
\cdot \colon G \times G \to G, \quad  (a,b) \mapsto a \cdot b \, .
\end{displaymath}
Die Verkn\"upfung muss
folgende Axiome erf\"ullen:}
\begin{itemize}
\item Assoziativit\"at: F\"ur alle Gruppenelemente $a, b$ und $c$ gilt: $(a \cdot b) \cdot c = a \cdot (b \cdot c).$
\item Es gibt ein neutrales Element $e \in G$, mit dem f\"ur alle Gruppenelemente $a \in G$ gilt:\\
$a \cdot e = e \cdot a = a.$
\item Zu jedem Gruppenelement $a \in G$ existiert ein inverses Element $a^{-1} \in G$ mit\\
$a \cdot a^{-1} = a^{-1} \cdot a = e.$
\end{itemize}
Aus den Gruppenaxiomen folgt von selbst, dass das neutrale Element eindeutig festgelegt ist, denn g\"abe es nebst einem
neutralen Element $n$ noch ein anderes $n' \neq n$, so g\"alte im Widerspruch dazu
$n \cdot n' = n = n'$. In der oben eingef\"uhrten, lediglich zweielementigen Symmetriegruppe 
des exakt spiegelsymmetrischen Tagpfauenauges \"ubernimmt $I$ die Rolle des neutralen Elements.

Die Verkn\"upfung zweier Gruppenelemente wird oft als Multiplikation durch einen Punkt dargestellt;
eine solche Notation ist praktisch, aber selbstverst\"andlich nicht zwingend.

Es bleibt zu bemerken, dass man im Falle des exakt spiegelsymmetrischen Tagpfauenauges eine
viel gr\"ossere Symmetriegruppe als die Menge $\{I,S\}$, die in der Mathematik auch als
Diedergruppe $D_1$ bezeichnet wird, zur Betrachtung heranziehen k\"onnte. Schliesslich w\"urde auch die
Vertauschung der beiden F\"uhler den Schmetterling nicht \"andern, ebenso der Austausch weiterer
beliebiger Teile des Schmetterlings, welche durch Spiegelung zur Deckung gebracht werden k\"onnen.
Es ist also in vielen F\"allen so, dass nicht die allgemeinsten denkbaren Symmetrietransformationen
f\"ur theoretische Untersuchungen herangezogen werden m\"ussen, sondern lediglich eine Auswahl derselben,
welche die Essenz der Symmetrieeigenschaften eines Objekts beschreiben.
Es muss f\"ur den Mathematiker zudem ganz klar erkl\"art sein, wie das Forschungsobjekt definiert ist.
F\"ur die bisherige Diskussion war es unwichtig, ob das Tagpfauenauge tats\"achlich ein r\"aumliches Wesen,
eingebettet im dreidimensionalen Raum darstellt oder ob sich die Betrachtungen lediglich auf das
zweidimensionale Bild des Tagpfauenauges bezogen, dieses eingebettet in einer zweidimensionalen Ebene
oder wiederum in einem dreidimensionalen Raum. Solche Spitzfindigkeiten sollen uns aber in der Folge
um der K\"urze willen nicht weiter aufhalten.

Weiter soll nicht unerw\"ahnt bleiben, dass die Spiegelung eines physikalischen Objekts als unphysikalische
Operation bezeichnet werden muss.
Die Spiegelung eines Standard-Menschen h\"atte zur Folge, dass dessen Herz durch massive chirurgische
Massnahmen von der linken auf die eher un\"ubliche
rechte Seite ger\"uckt werden m\"usste, und nicht nur dies. Sogar jedes einzelne Atom m\"usste gespiegelt werden.
Dennoch spielen in der theoretischen Physik Symmetrietransformationen, die sich lediglich als
Gedankenexperimente durchf\"uhren lassen, eine wichtige Rolle.
Eine Spiegelungstransformation wird in der Physik h\"aufig als \emph{Parit\"atstransformation} bezeichnet und durch
den Buchstaben $P$ bezeichnet. Weitere diskrete Transformationen von grosser theoretischer Relevanz
sind die \emph{Zeitumkehrtransformation} $T$ und die Umkehrung gewisser Teilcheneigenschaften wie der elektrischen
Ladung durch die sogenannte \emph{Ladungskonjugation} $C$.

Tats\"achlich sind die uns bekannten Naturgesetze unter keiner der eben erw\"ahnten
drei Transformationen exakt symmetrisch (Wu et al. 1957). Betrachtet man ein gespiegeltes Bild unserer Welt,
so laufen die entsprechenden Vorg\"ange nicht nach exakt denselben Gesetzen ab wie in unserer Welt.
Man nahm auch lange an, dass ein r\"uckw\"arts abgespielter Film eines physikalischen Vorgangs wiederum
einen realen physikalischen Vorgang zeigt. Dem ist aber nicht wirklich so. Die heute bekannte Verletzung der
Zeitumkehrsymmetrie der Naturgesetzte hat \"ubrigens nichts damit zu tun, dass uns die Vorg\"ange in einem
r\"uckw\"arts laufenden Film widernat\"urlich erscheinen. Es w\"are zwar befremdend, wenn sich eine zerbrochene
Tasse aus ihren Einzelteilen wieder zu einem Ganzen zusammensetzen w\"urde. Dennoch ist ein solcher Vorgang nicht
prinzipiell verboten, aber wegen der vorauszusetzenden Anfangsbedingungen sehr unwahrscheinlich.
In der Tat erh\"alt man aus einem r\"aumlich gespiegelt dargestellten realen
physikalischen Vorgang ($\rightarrow P$-Transformation, Parit\"atstransformation), 
bei dem zus\"atzlich alle Teilchen durch umgekehrt geladene ladungskonjugierte Varianten ersetzt wurden
($\rightarrow C$-Transformation, Ladungskonjugation) und welchen
man zeitumgekehrt ablaufen l\"asst ($\rightarrow T$-Transformation, Zeitumkehrtransformation),
wieder einen real existierenden physikalischen Prozess.
Diese sogenannte $CPT$-Invarianz der Naturgesetze gilt als fundamentale Symmetrie der
relativistischen Quantenfeldtheorie (L\"uders 1957). Eine Verletzung
der $CPT$-Invarianz w\"are ein wichtiger Hinweis auf neue Physik, konnte aber bis anhin nicht nachgewiesen
werden (D\"utsch und Gracia-Bondia 2012).

\subsubsection*{Kontinuierliche Symmetrien}
Um das Wesen der kontinuierlichen Symmetriegruppen zu veranschaulichen, betrachten wir als N\"achstes die
Symmetriegruppe einer Kugel im dreidimensionalen euklidischen Raum. Offensichtlich ist die Kugel
invariant unter beliebigen Drehungen (Rotationen) um ihren Mittelpunkt. Entsprechend ist die Gruppe aller m\"oglichen
Rotationen im dreidimensionalen Raum um den Mittelpunkt der Kugel eine Symmetriegruppe der Kugel. Diese Gruppe
wird in der Mathematik als spezielle orthogonale Gruppe in drei Dimensionen bezeichnet, kurz $SO(3)$.

Es existieren verschiedene M\"oglichkeiten, eine Rotation $R \in SO(3)$ zu charakterisieren. Naheliegend w\"are
beispielsweise die Angabe einer Drehachse und eines Drehwinkels. Diese beiden Gr\"ossen k\"onnen kompakt durch einen
einzelnen Vektor $\vec{\alpha}=(\alpha_1, \alpha_2 , \alpha_3)$ ausgedr\"uckt werden; der Betrag des Vektors
\begin{equation}
\alpha = | \vec{\alpha} | = \sqrt {\alpha_1^2 + \alpha_2^2 + \alpha_3^2}
\end{equation}
entspricht dann dem Drehwinkel der Rotation, der auf die L\"ange Eins normierte Einheitsvektor
\begin{equation}
\hat{\alpha}=\frac{\vec{\alpha}}{\alpha}
\end{equation}
charakterisiert die Richtung der Rotationsachse.
Es muss dabei aber ber\"ucksichtigt
werden, dass zwei solche Vektoren $\vec{\beta}=(\beta_1, \beta_2 , \beta_3)$ und 
$\vec{\gamma}=(\gamma_1, \gamma_2 , \gamma_3)$ dieselbe Rotation beschreiben, also
\emph{\"aquivalent} sind, wenn sie in dieselbe Richtung weisen und sich vom Betrag her um
ein ganzzahliges Vielfaches einer vollen Drehung um $360^o$ unterscheiden: Gilt
\begin{equation}
\vec{\beta}=r \cdot \vec{\gamma} \, \, \, \mbox{mit} \, \, \, r \in \mathds{R} \, \,  \, \mbox{und} \label{equiv}
\end{equation}
\begin{equation}
\beta-\gamma= n \cdot 360^o \, \, \, \mbox{mit} \, \, \, n \in \mathds{Z} \, , \label{equivalence}
\end{equation}
so sind die durch $\vec{\beta}$ und $\vec{\gamma}$ erkl\"arten Rotationen gleich:
\begin{equation}
(\ref{equiv},\ref{equivalence}) \, \, \Rightarrow \, \, R_{\vec{\beta}}= R_{\vec{\gamma}} \, .
\end{equation}
Die $SO(3)$ ist ein Beispiel f\"ur eine \emph{kontinuierliche Gruppe} oder eine \emph{Lie-Gruppe}.
Es ist n\"amlich m\"oglich, innerhalb der Gruppe selbst von einem Element zu einem anderen Element \"uber
einen zusammenh\"angenden, in gewissem Sinne kontinuierlichen Weg zu gelangen.
Zwischen zwei verschiedenen Rotationen $R_{\vec{\alpha}}$ und $R_{\vec{\beta}}$ liegen alle denkbaren
``Zwischenstufen'', und man gelangt von der Drehung $R_{\vec{\alpha}}$ kontinuierlich innerhalb der
Gruppe selbst zur Drehung $R_{\vec{\beta}}$, beispielsweise \"uber folgenden Weg im Raum $SO(3)$ der
Drehungen
\begin{equation}
R_{(1-s) \vec{\alpha} + s \vec{\beta}} \, , \, \, \, s \in [0,1] \, , 
\end{equation}
wobei der Parameter $s$ kontinuierlich von $0$ bis $1$ variiert wird.

Im Falle der Diedergruppe $D_1$, der Symmetriegruppe des perfekten Tagpfauenauges, ist eine solche kontinuierliche
Variation nicht m\"oglich. Es gibt keine ``halbe Spiegelung'', aber sehr wohl eine Rotation um einen halben
Winkel im Falle der $SO(3)$ bei vorgegebenem Drehwinkel um eine vorgegebene Drehachse.

Die allgemeing\"ultige Definition kontinuierlicher Gruppen beruht auf Begriffen, welche im
mathematischen Teilgebiet der \emph{Topologie} definiert sind. Die obige Diskussion enth\"alt aber
die wesentlichen intuitiven Aspekte.

Man k\"onnte nun meinen, dass die Drehgruppe $SO(3)$ tats\"achlich eine volle Symmetriegruppe
unserer Naturgesetze sein sollte, da zumindest lokal erfahrungsgem\"ass jedes physikalische
System bei Vernachl\"assigung \"ausserer, nicht notwendigerweise rotationssymmetrischer
Einfl\"usse wie dem Schwerefeld der Erde im Labor nach genau denselben
Regeln funktioniert wie ein analoges, aber gedrehtes System. Dem ist aber nicht so, und diese
im folgenden Abschnitt diskutierte tiefgreifende Eigenschaft der Natur erlaubt eine
Klassifikation der uns bekannten Elementarteilchen in zwei Klassen.

\subsection*{Symmetrien in der Teilchenphysik}
\subsubsection*{Bosonen und Fermionen}
In der Physik erfordert die Durchf\"uhrung eines Experiments die Trennung  der Welt in einen
Beobachter (oder Experimentator) und eine Versuchsanordnung (oder ein zu untersuchendes physikalisches
System). Es soll hier nicht auf die philosophischen Aspekte
eingegangen werden, die durch einen solchen gedanklichen Vorgang aufgeworfen werden.
Die Frage aber, um welchen Winkel ein physikalisches System um eine beliebige feste Achse
gedreht werden muss, bis es wieder in
derselben Beziehung zu seiner Umwelt steht und somit der urspr\"ungliche Zustand des Gesamtsystems
erreicht ist, hat sich als sehr fruchtbar erwiesen.

Stellen wir uns also vor, dass ein einfaches physikalisches Objekt, beispielsweise ein Proton,
aktiv gedreht wird, wobei das Proton ansonsten im Wesentlichen unbeeinflusst bleiben soll.
Die naheliegende Annahme, dass nach einer vollen Umdrehung um $360^o$ der Urzustand des
Gesamtsystems \emph{Kosmos - Proton} wieder hergestellt sei, ist
falsch! Ein einfaches Alltags-Experiment, dargestellt in den Abbildungen 3 und 4, veranschaulicht
diese Tatsache. Dreht man seine Hand um $360^o$ um eine feste senkrechte Achse wie in den Bildern
dargestellt, so wird dadurch der ungedrehte Urzustand offensichtlich nicht wieder erreicht; die
resultierende Gesamtsituation des Systems \emph{Mensch - Hand} ist zwar frei von physischen
Verletzungen, aber gespannt. Erstaunlicherweise
kann der Urzustand wieder hergestellt werden, wenn die Hand um weitere $360^o$ in derselben
Drehrichtung rotiert wird, insgesamt also um $720^o$!

}
\end{multicols}

\begin{center}
\noindent \includegraphics[width=2.9cm,height=4.1cm]{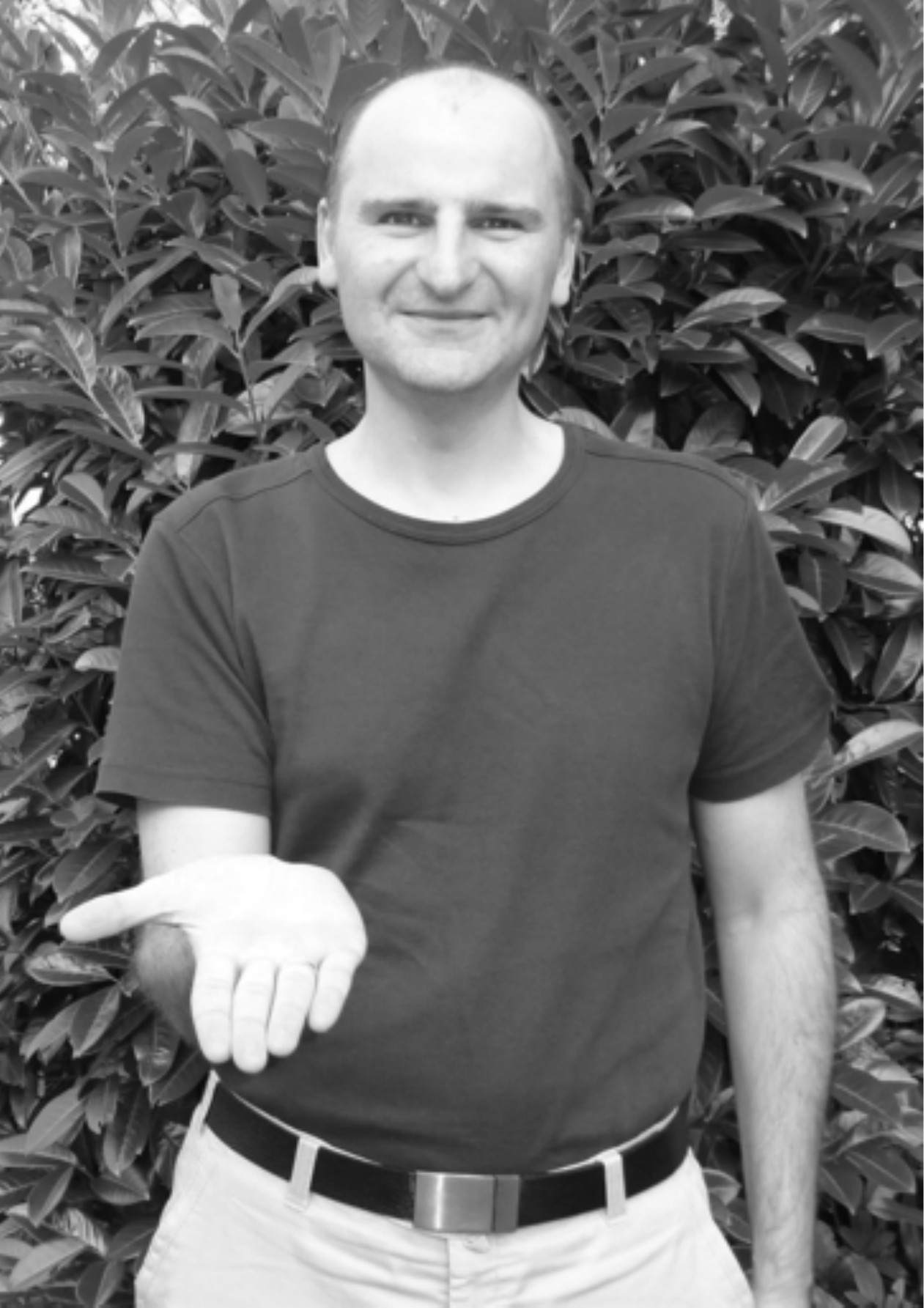}
\includegraphics[width=2.9cm,height=4.1cm]{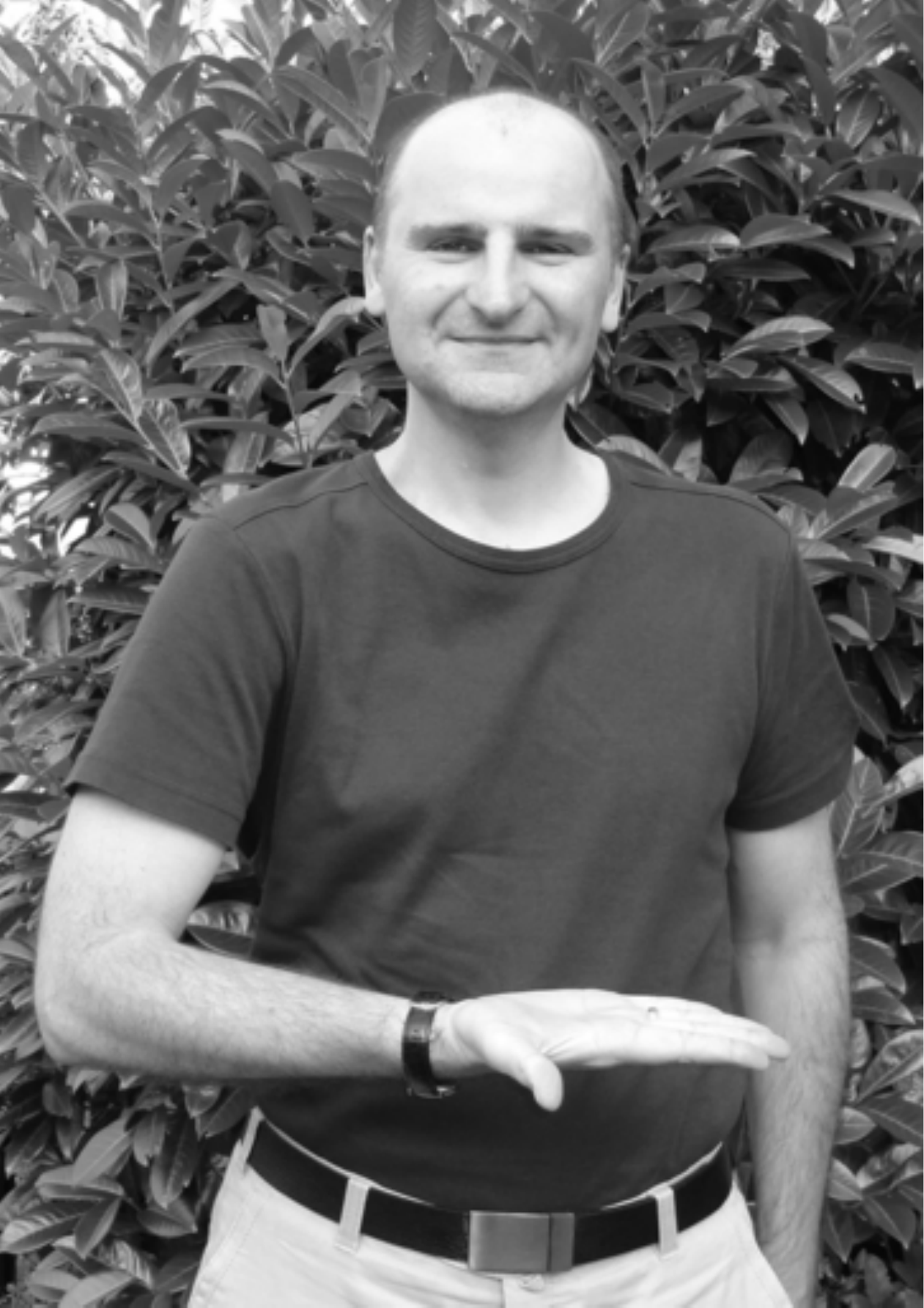}
\includegraphics[width=2.9cm,height=4.1cm]{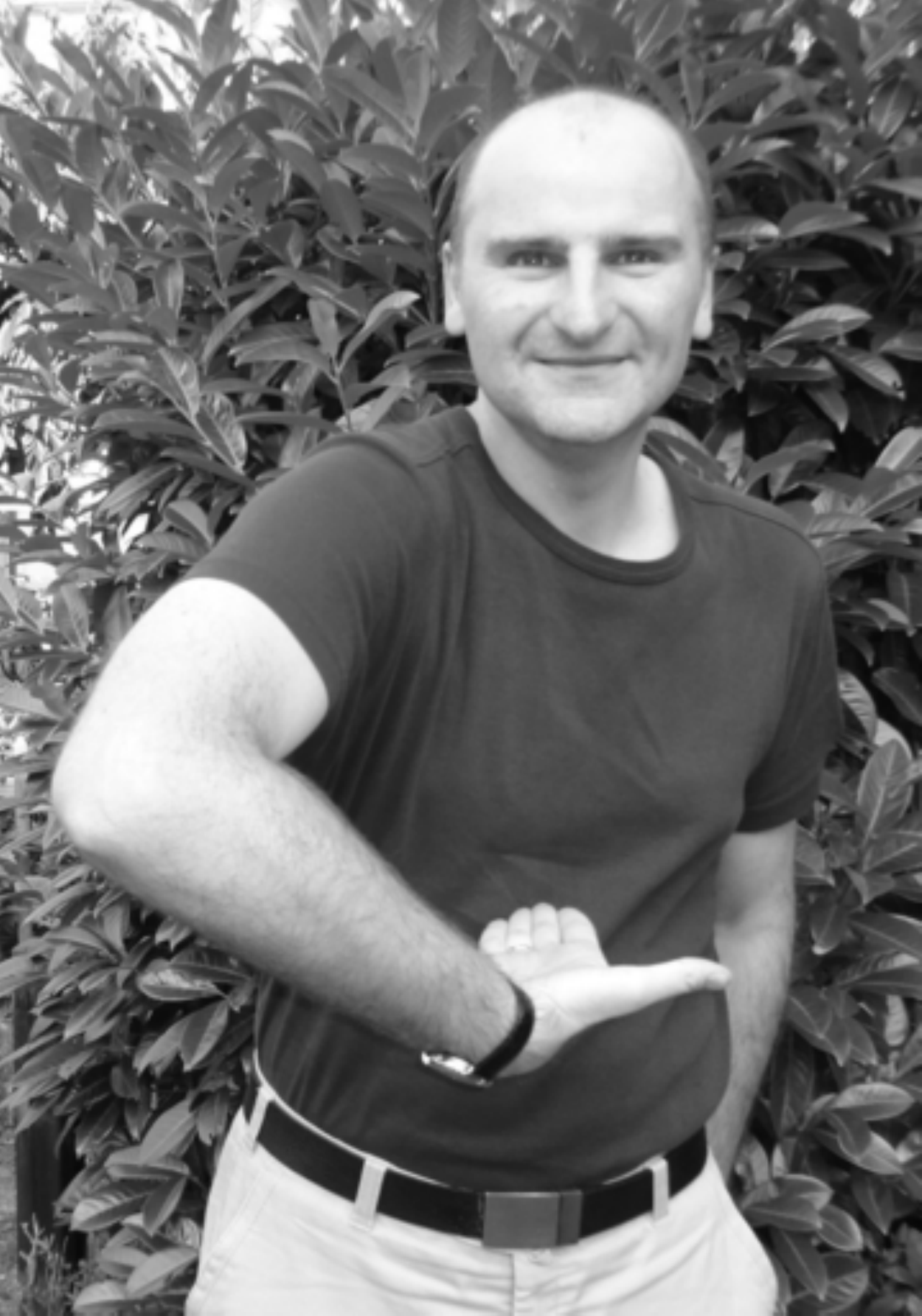}
\includegraphics[width=2.9cm,height=4.1cm]{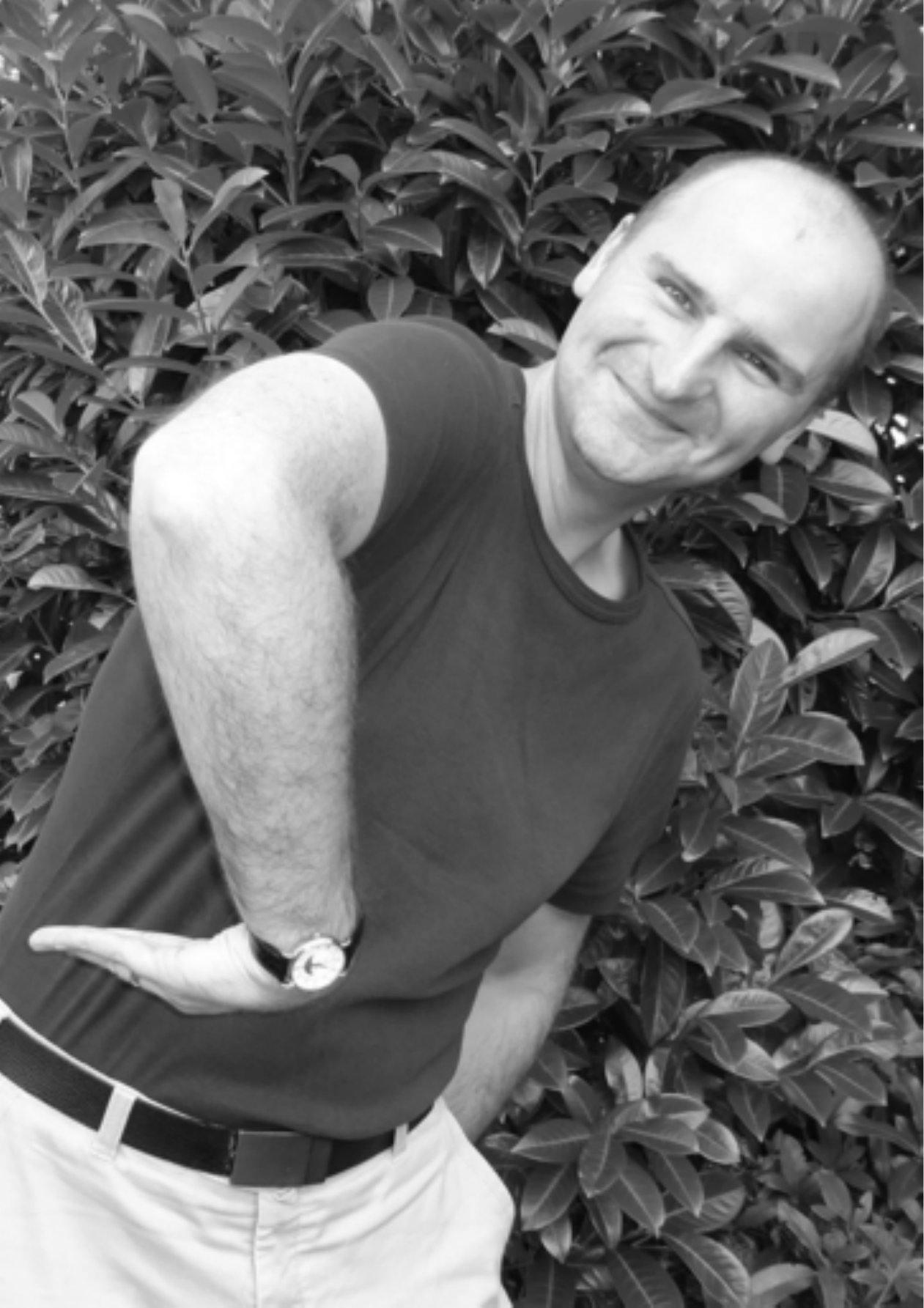}
\includegraphics[width=2.9cm,height=4.1cm]{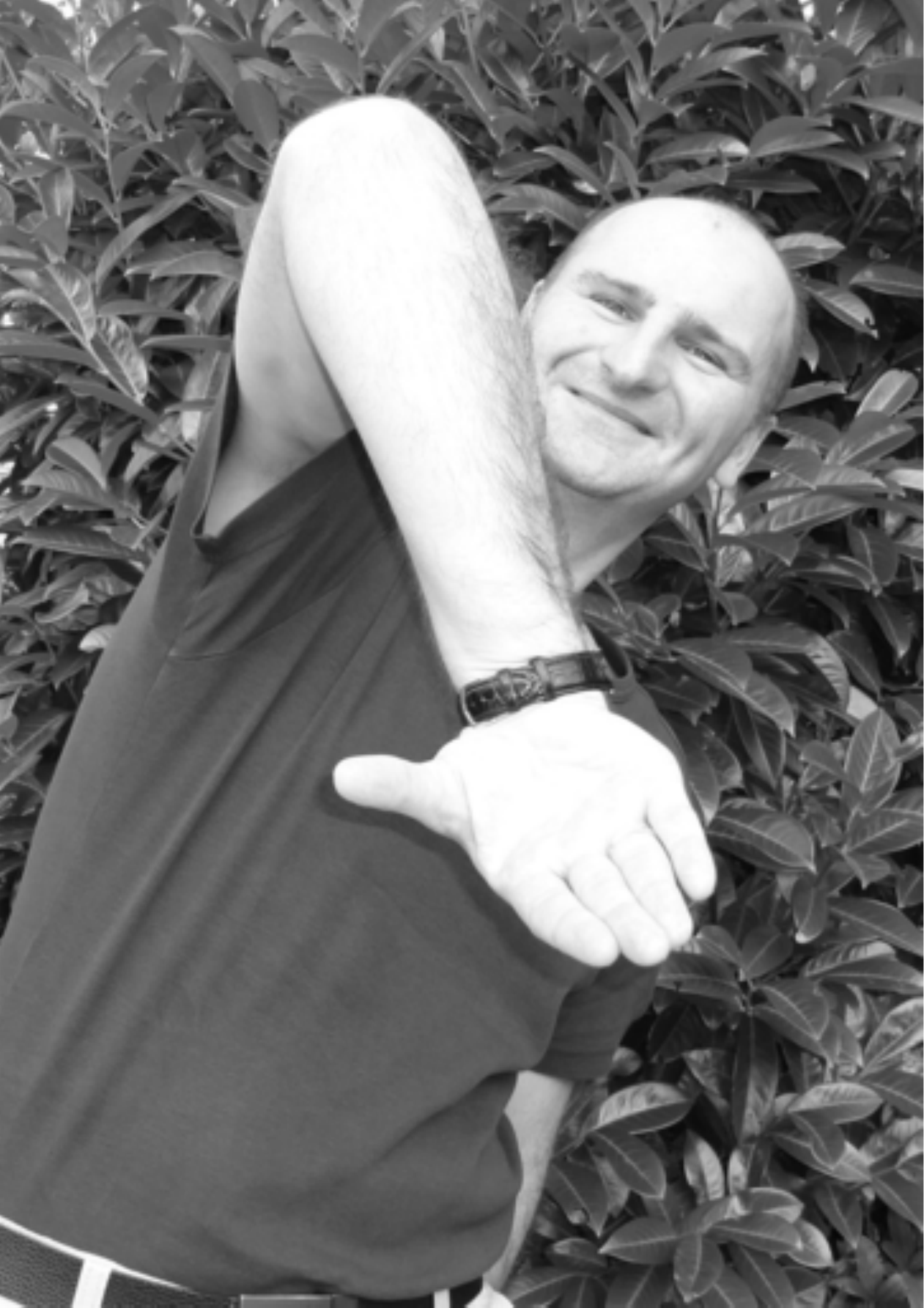}
\end{center}
\noindent  {\small{{{\bfseries{Abb. 3:}}
Ein um einen Winkel von 360$^o$ gedrehtes Objekt als Teil des Ganzen
steht nicht notwendigerweise in derselben Beziehung zum Kosmos wie
vor der Drehung, wie der oben in Viertelsdrehungen dargestellte Selbstversuch zeigt. Die Rotation einer Hand um
eine senkrechte Drehachse resultiert nach Abschluss der vollen Umdrehung in einer f\"ur den Experimentator
eher ungem\"utlichen Endsituation, die sich von der entspannten Ausgangslage offensichtlich unterscheidet
(Fotos: Irene Aste).}}}
\normalfont
\begin{center}
\noindent \includegraphics[width=2.9cm,height=4.1cm]{Bild5gg.pdf}
\includegraphics[width=2.9cm,height=4.1cm]{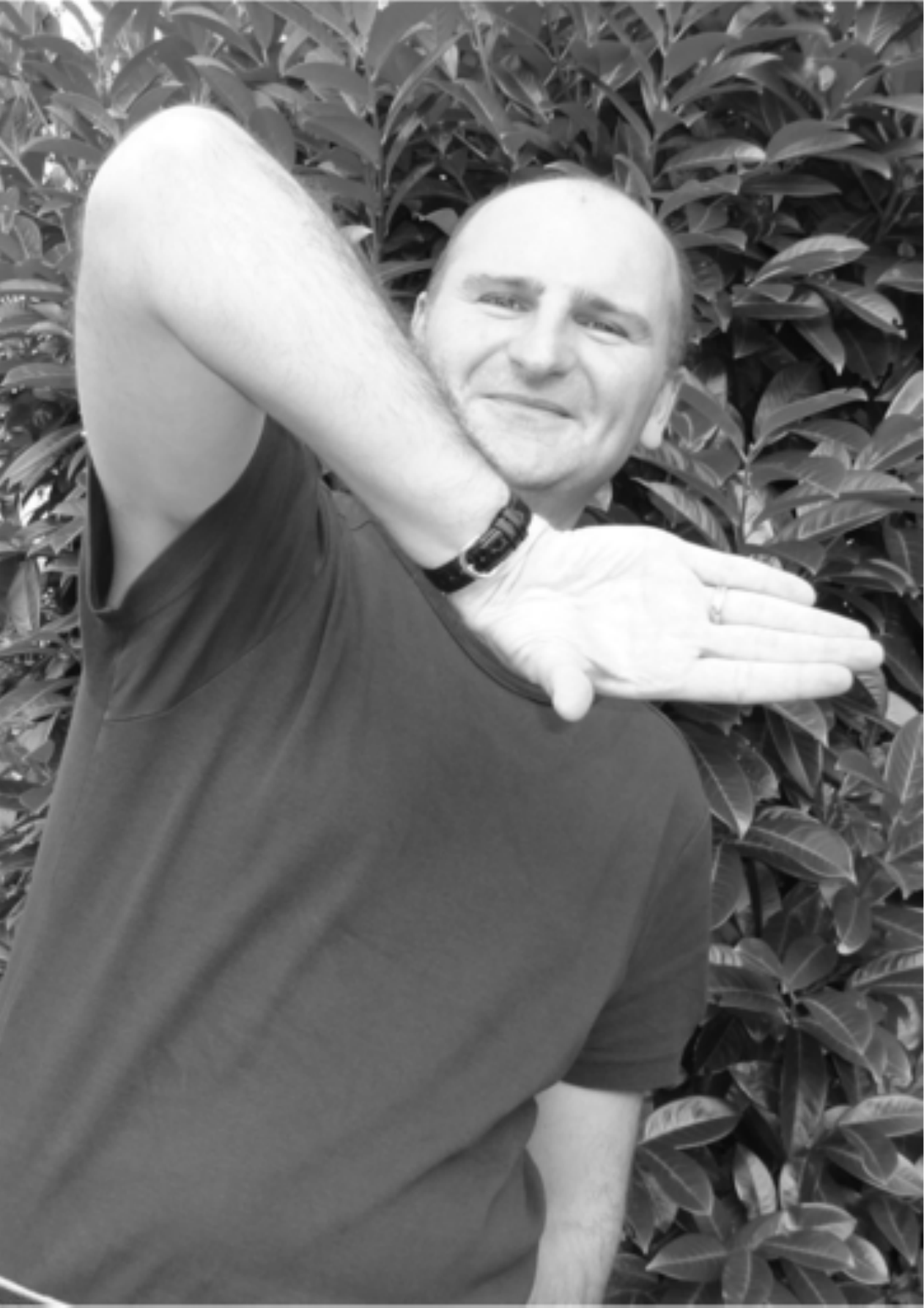}
\includegraphics[width=2.9cm,height=4.1cm]{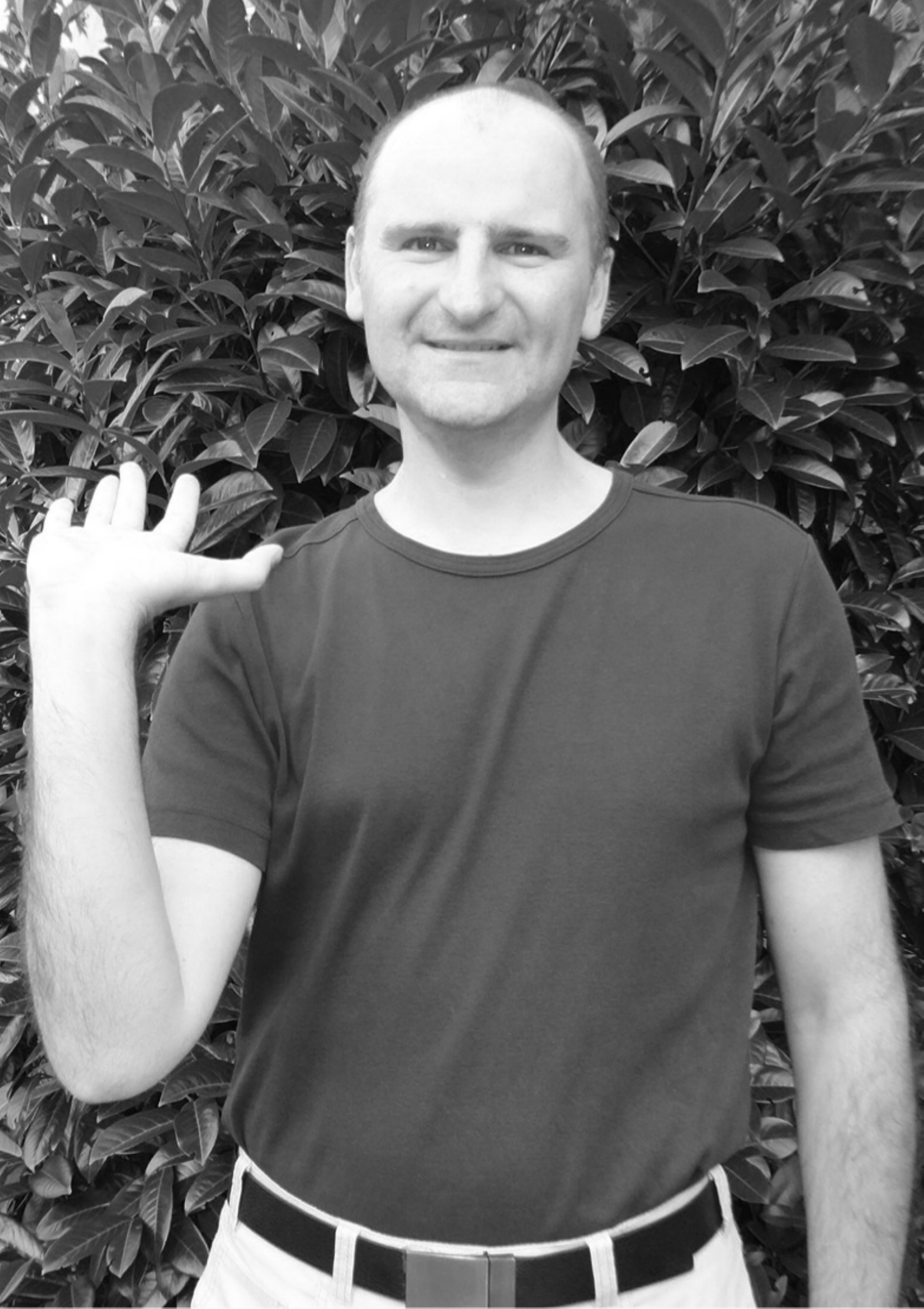}
\includegraphics[width=2.9cm,height=4.1cm]{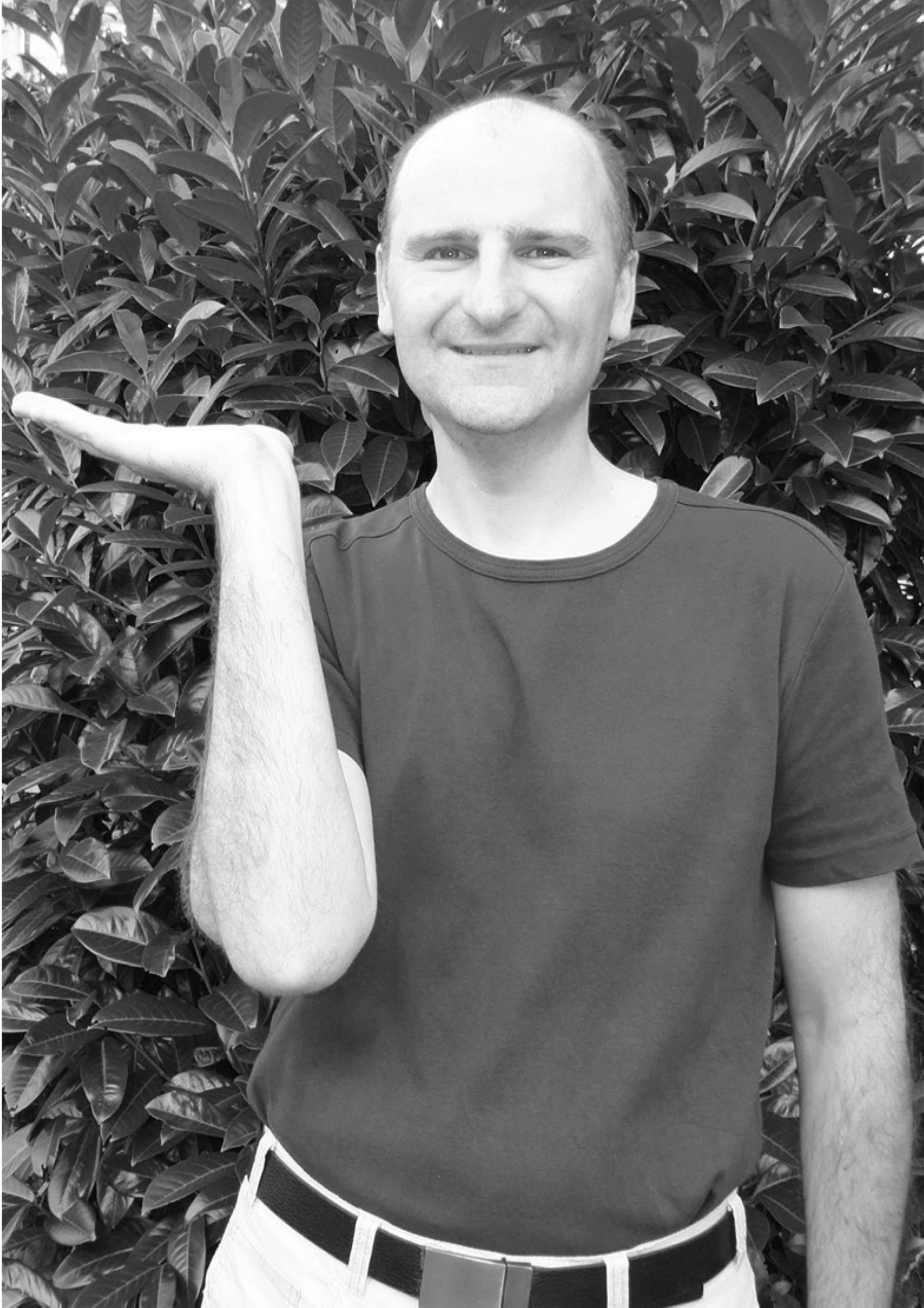}
\includegraphics[width=2.9cm,height=4.1cm]{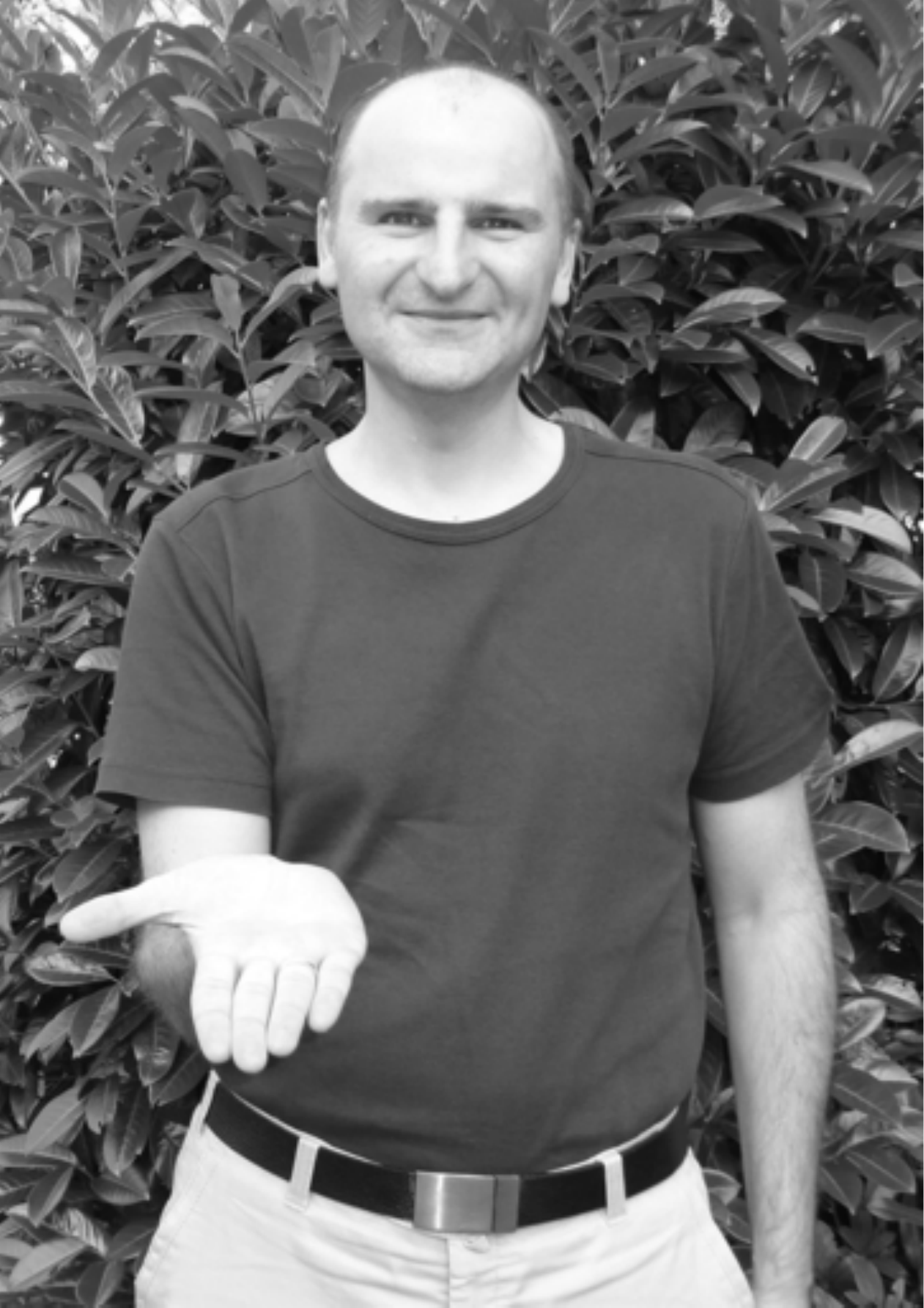}
\end{center}
\noindent {\small{{{\bfseries{Abb.~4:}} Eine verbl\"uffende Situation entsteht, wenn die Hand aus {\bfseries{Abb. 3}}
im {\itshape{urspr\"unglichen Drehsinn}} weiter gedreht wird,
bis ein Drehwinkel von $720^o$ erreicht ist.
Tats\"achlich stellt sich erst nach zwei vollen Umdrehungen wieder der Urzustand des Mensch-Hand-Systems ein
(Fotos: Irene Aste).}}}

\begin{multicols}{2}{
Es hat sich herausgestellt, dass die in der Natur vorkommenden Elementarteilchen strikt in zwei Klassen einteilbar
sind: Die eine Klasse sind die \emph{Bosonen}, so benannt zu Ehren des indischen Physikers Satyendranath Bose.
In Falle der Bosonen ist bereits nach einer vollen Umdrehung wieder der relative Urzustand zum Kosmos
hergestellt. Die zu Ehren des Physik-Nobelpreistr\"agers Enrico Fermi als
Fermionen bezeichneten Mitglieder der anderen Klasse hingegen
bed\"urfen zur Wiederherstellung des Urzustands tats\"achlich einer Drehung um $720^o$.
Jeder Teil unserer Welt ist mit dieser in unaufl\"oslicher Weise verwoben, im Falle der Fermionen
in einer f\"ur die allt\"agliche Erfahrung verwirrenden Weise.

Ein Teilsystem kann somit niemals in Isolation vom Rest der Welt betrachtet werden.
Die Erkenntnis, dass Elektronen tats\"achlich Fermionen sind, hat die Erkl\"arung der Atomstruktur
erst m\"oglich gemacht (Pauli 1925). Bosonen und Fermionen zeigen bei kollektivem Auftreten
sehr unterschiedliche Verhaltensweisen. In etwas oberfl\"achlicher und qualitativer Manier kann behauptet werden, dass
Fermionen als Materiebausteine aufgefasst werden d\"urfen, w\"ahrend Bosonen f\"ur den Aufbau
von Kraftfeldern zwischen der fermionischen Materie verantwortlich sind.
Zu den Fermionen geh\"oren im aktuell g\"ultigen Standard-Modell (SM) der Elementarteilchenphysik die Quarks
und die Leptonen, zu welchen das Elektron, das Myon und das Tauon geh\"oren, wie auch die Neutrinos.
Zu den Bosonen z\"ahlen das Higgs-Teilchen sowie das Photon, welches die elektromagnetischen Kr\"afte vermittelt,
die elektrisch geladenen W$^+$- und W$^-$-Bosonen und das neutrale Z-Boson als Vermittler der schwachen
Kernkraft und schliesslich die Gluonen, welche f\"ur die starken Kernkr\"afte zwischen den Quarks
verantwortlich gemacht werden.
Vernachl\"assigt man gravitative Effekte, so ist diese Liste der fundamentalen Teilchen im SM vollst\"andig.
Die Gravitation l\"asst sich durch die naive Einf\"uhrung eines bosonischen Teilchens, dem \emph{Graviton},
beschreiben. Allerdings sind die damit verbundenen
mathematischen Komplikationen schwerwiegend und bis dato eigentlich unverstanden (Aste et al. 2009).

Die Unterteilung der Elementarteilchen in Bosonen und Fermionen ist universell und spielt
dieselbe Rolle in hypothetischen Theorien wie dem \emph{minimalen supersymmetrischen Standardmodell (MSSM)},
einer popul\"aren Erweiterung des SM. Sie spiegelt die Tatsache wieder, dass die ``naive''
Rotationsgruppe $SO(3)$ nicht die eigentliche Symmetriegruppe der Natur darstellt,
zumindest in einer $3+1$-dimensionalen Raum-Zeit. Die $SO(3)$ wird daher
im mathematischen Formalismus der Quantenmechanik ersetzt durch ihre sogennante doppelte
\"Uberlagerungsgruppe $SU(2)$, welche die in den Abbildungen 3 und 4 dargestellte doppelschichtige Natur
der Drehungen zu beschreiben vermag. Diese wichtige Gruppe der speziellen unit\"aren Transformationen
in zwei komplexen Dimensionen $SU(2)$, auf die wir an dieser Stelle nicht weiter eingehen
wollen, beschreibt in den modernen Theorien der Elementarteilchenphysik auch so genannte
\emph{innere Symmetrien}, also Symmetrien, die nicht ``\"ausserer'' r\"aumlicher Natur sind wie
die Rotationssymmetrie, sondern welche mit inneren Qualit\"aten der Teilchen wie beispielsweise
Ladungszust\"anden verkn\"upft sind.

Der obigen Diskussion muss die Bemerkung angef\"ugt werden, dass die rein r\"aumlich motivierte und doch
im Wesentlichen korrekt begr\"undete Unterscheidung
von Fermionen und Bosonen durch die Gruppentheorie der $SO(3)$ und $SU(2)$ eigentlich in einem erweiterten,
die spezielle Relativit\"atstheorie ber\"ucksichtigenden Rahmen untersucht werden muss.

\subsubsection*{Materie und Antimaterie}
Eine weitere, bereits anget\"onte Symmetrie ist die sogenannte $CPT$-Symmetrie, welche besagt,
dass ein ladungskonjugierter, r\"aumlich und zeitlich gespiegelter physikalischer Prozess wiederum einem
m\"oglichen physikalischen Prozess entspricht. Die sogenannte $C$-Konjugation oder Ladungskonjugation
ist dabei eine Transformation, die angewendet auf einen Teilchenzustand gewisse Eigenschaften wie die Masse des Teilchens
exakt gleich l\"asst, andere Eigenschaften aber \"andert. So \"andert die Ladungskonjugation das Vorzeichen
der elektrischen Ladung eines Teilchens. Es ist aber streng genommen nicht so, dass ein $C$-konjugierter
physikalischer Zustand tats\"achlich wieder einem physikalischen Zustand entspricht. Erst wenn der $C$-konjugierte
Zustand zus\"atzlich einer r\"aumlichen Spiegelung $P$ und der Zeitspiegelung $T$ unterworfen wird, liegt wieder ein
physikalischer Zustand mit gegebenenfalls neuen Eigenschaften vor. Die Naturgesetze besitzen keine
exakte Symmetrie unter den Transformationen $C$, $P$ und $T$.

Streng genommen existiert daher zu jedem Teilchentyp ein Antiteilchentyp, der durch die drei Operationen
$CPT$ aus der urspr\"unglichen Teilchensorte hervorgeht.
Immer gilt, dass sowohl Teilchen wie auch Antiteilchen dieselbe Masse und
die umgekehrte elektrische Ladung besitzen. Zugleich ist es bei elektrisch neutralen Teilchen
m\"oglich, dass Teilchen ihren Antiteilchen entsprechen. So repr\"asentieren zwar die elektrisch ungeladenen Neutronen
eine andere Sorte von Teilchen als die Antineutronen, Photonen sind aber in gewissem Sinne ihre eigenen Antiteilchen.
Im Falle der elektrisch neutralen Neutrinos ist es tats\"achlich noch nicht restlos gekl\"art,
wie sie sich unter den $C$- $P$- und $T$-Transformationen verhalten.

Astronomische Beobachtungen legen nahe, dass in unserem Universum ein \"Uberschuss
an gew\"ohnlicher Materie gegen\"uber der Antimaterie herrscht. Langj\"ahrige astronomische Beobachtungen
lassen es sehr unwahrscheinlich erscheinen, dass es in unserem Universum gr\"ossere Ansammlungen von
Antimaterie gibt (Canetti et al. 2012). Offensichtlich wurde beim Urknall eine gr\"ossere Menge Materie
als Antimaterie erzeugt, sodass nach der gegenseitigen Ausl\"oschung der beiden Materiesorten ein
\"Uberschuss an Materie \"ubrig blieb. Theoretische Untersuchung suchen nach m\"oglichen Gr\"unden f\"ur diese nicht
ganz gelungene Ausl\"oschung in einer Verletzung der $CP$-Symmetrie, der wir unsere Existenz verdanken.
Die vorl\"aufigen Resultate sind aber noch nicht schl\"ussig.

Nebst der Teilchenklassifikation in Bosonen und Fermionen vermittels des Transformationsverhaltens derselben
unter der speziellen unit\"aren Gruppe in zwei komplexen Dimensionen $SU(2)$ erlaubt also die diskrete
$CPT$-Symmetrie eine weitere Unterscheidung von Teilchen und ihren $CPT$-konjugierten Zust\"anden, den
Antiteilchen.

\subsubsection*{Spontane Symmetriebrechung}
Eine ganz allt\"gliche, aber vielen Leuten g\"anzlich unbewusste Tatsache
beruht auf der Beobachtung, dass die Symmetrie eines Naturgesetzes im Allgemeinen
nicht der Symmetrie der Objekte gleich ist, welche dem Gesetz unterworfen sind. 
Dies l\"asst sich durch eine einfache Aufgabe leicht veranschaulichen.
Man stelle sich vor, dass vier St\"adte aus Spargr\"unden durch ein m\"oglichst kurzes Bahnnetz miteinander verbunden
werden sollen. Zuf\"alligerweise sollen die Standorte der Bahnh\"ofe dieser St\"adte alle pr\"azise auf den
Eckpunkten eines Quadrates liegen. Vereinfachend nehmen wir an, dass die
Kantenl\"ange dieses Quadrats gerade eine Masseinheit betr\"agt, welche wir nicht weiter
notieren wollen. Jede Stadt besitzt also zwei Nachbarst\"adte im Abstand $1$, und eine weitere
Nachbarstadt im Abstand $\sqrt{2} \simeq 1.4142 \ldots$ .

Naiverweise w\"urde man erwarten, dass die L\"osung des Problems dieselbe Symmetrie aufweisen sollte
wie das Quadrat. Auf diese L\"osung wollen wir nun kurz eingehen.
Ein Quadrat ist sicher invariant unter folgenden Transformationen: der Identit\"at $I=R_0$, welche als Rotation
um einen Winkel von $0^o$ aufgefasst werden kann, sowie unter Drehungen $R_{1/4}$, $R_{1/2}$,
$R_{3/4}$ um $90^o$, $180^o$ und $270^o$ im mathematisch positiven Sinne, dem Gegenuhrzeigersinn.
Hinzu kommen Spiegelungen an Achsen, wie sie
in Abb.~5 eingezeichnet sind. Diese wollen wir mit $S_0$ (horizontale Achse), $S_{1}$ (im Gegenuhrzeigersinn
gegen\"uber der horizontalen Achse um $45^o$ gedrehte Achse), $S_{2}$ (vertikale Achse) sowie mit
$S_3$ f\"ur die verbleibende Achse bezeichnen. Nat\"urlich liessen sich die Spiegelungen auch als
r\"aumliche dreidimensionale Drehungen von $180^o$ um die Spiegelachsen auffassen. 

Die Symmetriegruppe unseres Problems besteht also aus den Elementen
\begin{equation}
\{ R_0, \, R_{1/4}, \, R_{1/2}, \, R_{3/4}, \, S_0, \, S_{1}, \, S_{2} , \, S_3  \} \, .
\end{equation}
Die Gruppenelemente lassen sich verkn\"upfen oder ``multiplizieren'', es gilt beispielsweise
\begin{equation}
R_{1/2} \cdot R_{3/4} = R_{3/4} \cdot R_{1/2} = R_{1/4} \, ,
\end{equation}
denn eine Drehung um $270^o$ gefolgt von einer Drehung um $180^o$ \emph{oder umgekehrt} resultiert letztlich in
einer Drehung um $90^o$,
oder
\begin{equation}
S_1 \cdot S_3 = R_{1/2} \, .
\end{equation}
Erstaunlicherweise gilt aber (die rechte Transformation in einem Produkt wird per Abmachung
zuerst ausgef\"uhrt)
\begin{equation}
S_1 \cdot R_{1/4} = S_0 \, ,  \quad R_{1/4} \cdot S_1 = S_2 \, ,
\end{equation}
es ist also
$S_1 \cdot R_{1/4} \neq R_{1/4} \cdot S_1 \,$. Probieren Sie es selbst mit einem beschrifteten
Papierquadrat aus!

Anders als bei der von den Verkn\"upfungen Addition und Multiplikation reeller Zahlen her bekannten Situation
gilt bei Drehungen und Spiegelungen also, dass die Reihenfolge ihrer Verkn\"upfung eine Rolle
spielt, es ist zwar $3+4=4+3$ und $3 \cdot 4 = 4 \cdot 3$, doch Drehungen reagieren empfindlich auf
Vertauschung.

Gruppen, bei denen die Verkn\"upfungsreihenfolge ihrer Elemente keine Rolle spielt,
heissen zu Ehren des norwegischen Mathematikers Nils Henrik Abel \emph{abelsch}, andernfalls entsprechend
\emph{nicht-abelsch}. Die nicht-abelsche Symmetriegruppe unseres Vier-St\"adte-Problems wird
in der Mathematik als Diedergruppe $D_4$ bezeichnet.

\begin{center}
\includegraphics[width=4.77cm,angle=90]{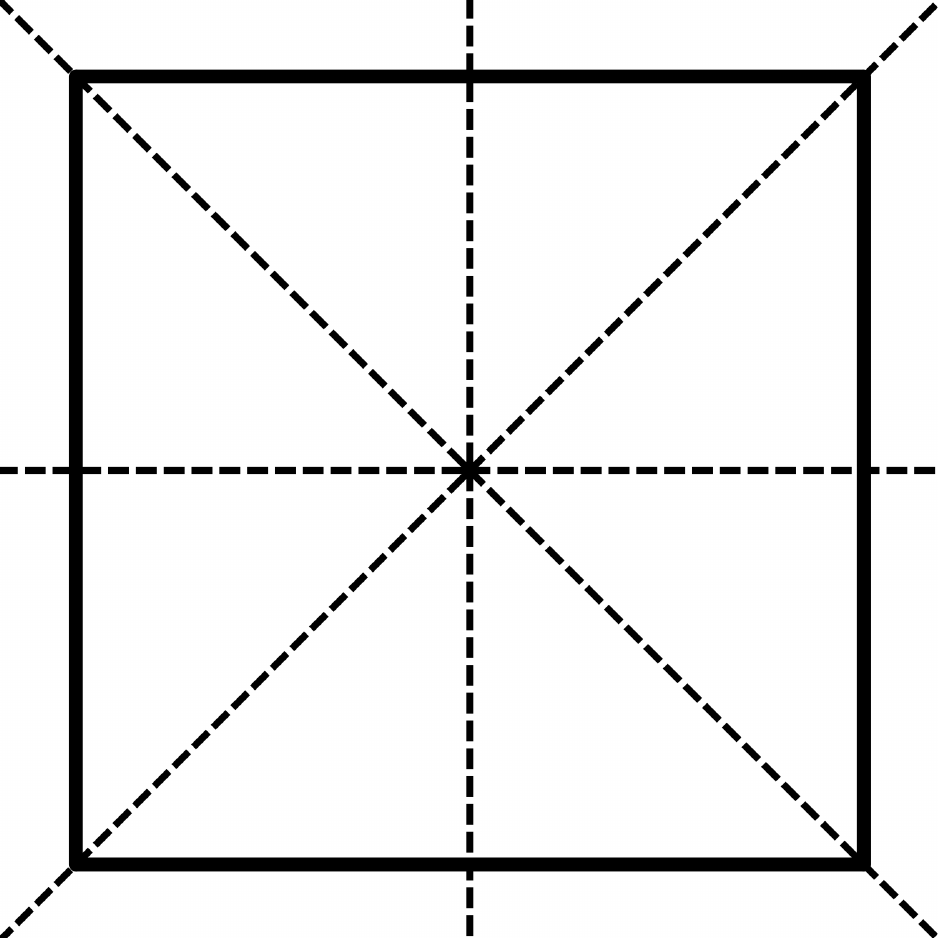}
\end{center}
\noindent  {\small{{\bfseries{Abb.~5:}} Achsen der Spiegelungen, welche das dargestellte Quadrat wieder
zur Deckung bringen.}}

\begin{center}
\includegraphics[width=4.2cm,angle=90]{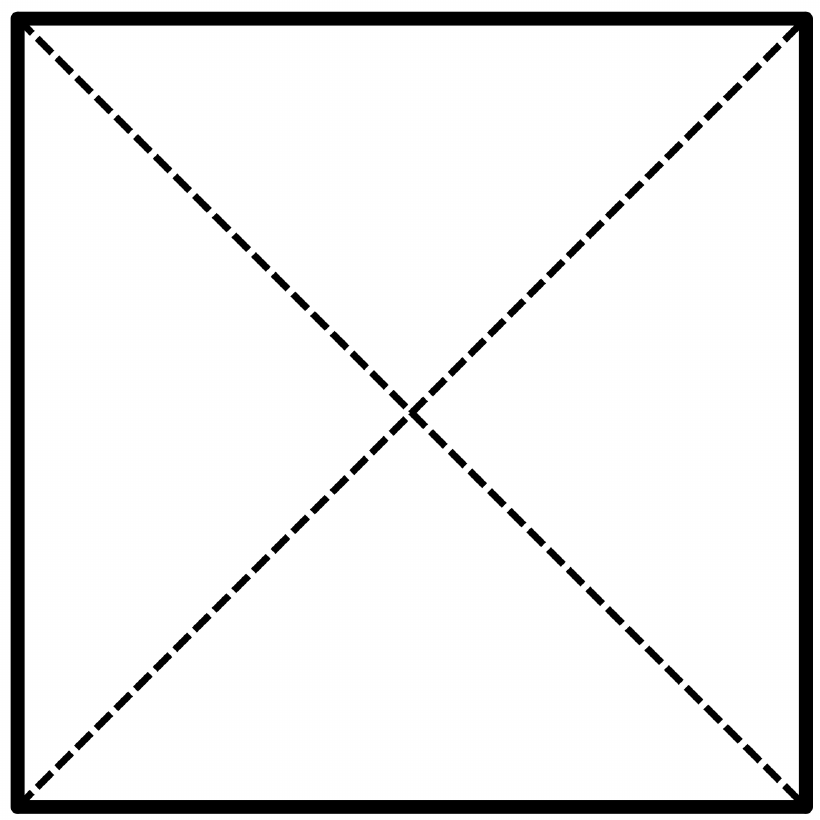}
\end{center}
\noindent {\small{{\bfseries{Abb.~6:}} Verbindungsnetz (gestrichelte Strecken) der Gesamtl\"ange $\sqrt{8} \simeq
2.8284 \ldots$ zwischen den Eckpunkten eines Einheitsquadrates der Kantenl\"ange 1. Das Verbindungsnetz
besitzt zum Quadrat analoge Symmetrieeigenschaften, ist aber nicht optimal.}}\\

Wie aber sieht nun das k\"urzeste Verbindungsnetz zwischen den Eckpunkten eines Quadrates aus?
Nat\"urlich bilden die Kanten des Quadrats ein Verbindungsnetz der L\"ange $4$ zwischen allen vier St\"adten.
N\"utzt man aber die Diagonalen im Quadrat wie in Abb.~6 dargestellt aus, so ist jede Stadt von einer anderen
Stadt aus \"uber ein k\"urzeres Verbindungsnetz der Gesamtl\"ange $\sqrt{8}=2 \sqrt{2} \simeq 2.8284 \ldots$
zu erreichen. Beide bisher pr\"asentierten Verbindungsnetzvorschl\"age besitzen dieselbe Symmetriegruppe wie das Quadrat,
sind invariant unter den Drehungen und Spiegelungen der Diedergruppe $D_4$. Es geht aber noch besser.

Abb.~7 zeigt zwei Verbindungsnetze der Gesamtl\"ange $\sqrt{3}+1 \simeq 2.7320 \ldots $,
welche tats\"achlich $3.41 \%$ k\"urzer sind als der in Abb.~6 dargestellte Vorschlag der L\"ange $\sqrt{8} \simeq
2.8284 \ldots$. Beide im Wesentlichen gleichwertigen und tats\"achlich \emph{optimalen}
L\"osungen besitzen eine kleinere Symmetriegruppe
als das gestellte Problem! Im Gegensatz zum Quadrat kommen sie bei einer Drehung um $90^o$ nicht
mit sich selbst zur Deckung. Die Symmetriegruppe der L\"osungen ist eine \emph{Untergruppe} der $D_4$
und besteht aus den vier Transformationen
\begin{equation}
D_2 = \{ R_0, \, R_{1/2}, \, S_0, \, S_{2}  \} \, .
\end{equation}
\begin{center}
\includegraphics[width=9.05cm,angle=90]{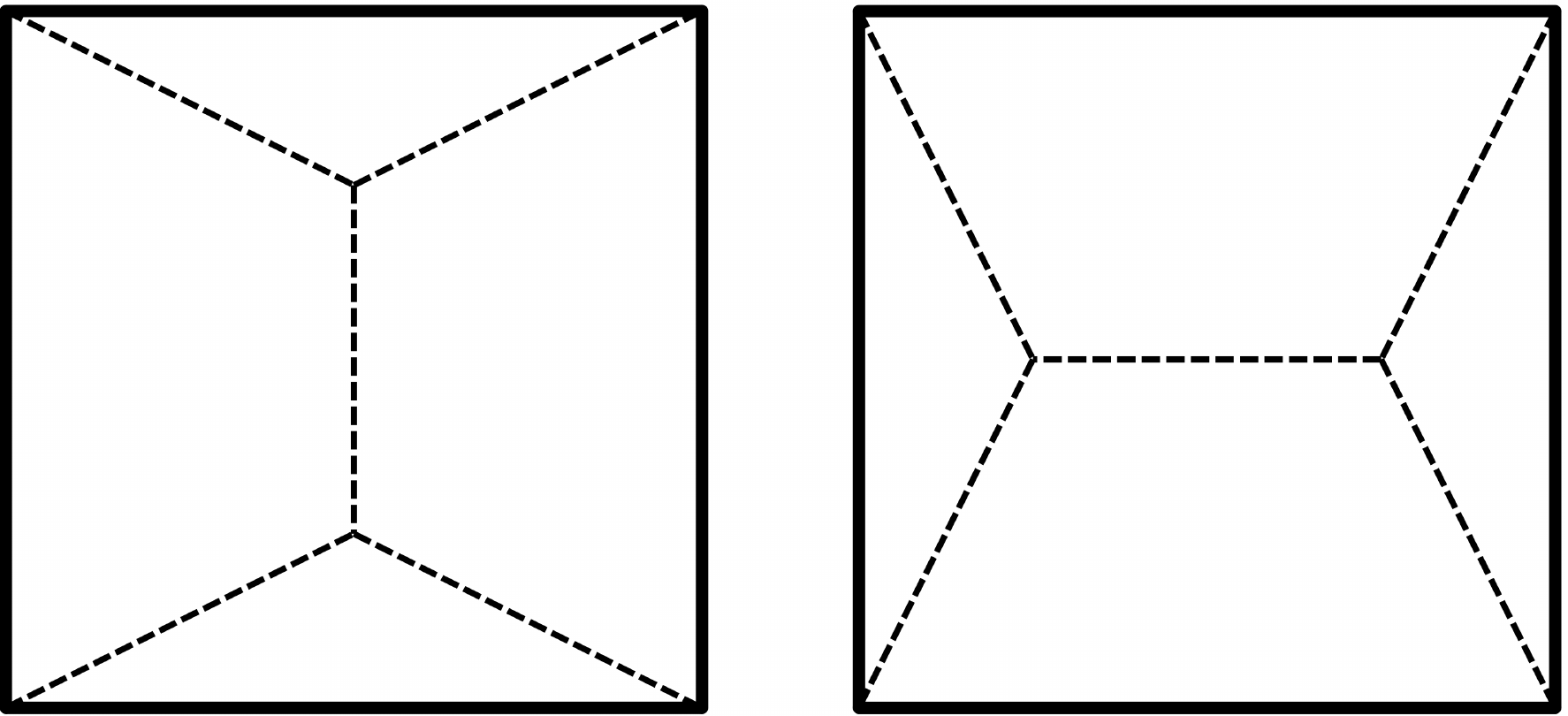}
\end{center}
\noindent  {\small{{{\bfseries{Abb.~7:}} K\"urzeste Verbindungsnetze (gestrichelte Strecken)
der Gesamtl\"ange $\sqrt{3}+1\simeq 2.7320 \ldots $
zwischen den Eckpunkten eines Einheitsquadrates der Kantenl\"ange 1. Die beiden kleinen Winkel in den Dreiecken betragen $30^o$.}}}\\

Scherzhafterweise k\"onnte man hier anbringen, dass die spontane Symmetriebrechung eine Erkl\"arung f\"ur den
S\"undenfall liefert: Die g\"ottlichen Gesetzte waren zwar wohlgeordnet und von h\"ochster Perfektion,
doch bei der Ausf\"uhrung der Gesetzte haperte es dann schliesslich doch.

Die spontane Symmetriebrechung ist es letztlich, welche die Existenz einer nicht trivialen
Welt erm\"oglicht. Die Naturgesetze selbst weisen in zumindest sehr guter N\"aherung eine Translations- und
eine Rotationssymmetrie auf. So sind die Naturgesetze in Basel dieselben wie hinter dem Mond.
Ebenso funktioniert ein Taschenrechner genau gleich, wenn er um einen beliebigen Winkel
gedreht wird, um es erneut plakativ auszudr\"ucken. Dennoch besteht die Welt nicht aus einem
homogenen Medium, sondern aus nicht rotationssymmetrischen Entit\"aten wie St\"uhlen, Weinflaschen
und Singv\"ogeln, die durch Drehung und Verschiebung in einen ver\"anderten Zustand versetzt werden, doch nicht die Naturgesetze,
denen die Dinge unterliegen.

Theoretische Modelle mit spontaner Symmetriebrechung spielen in der Elementarteilchenphysik und vielen
anderen Disziplinen der Physik eine wichtige Rolle
im Zusammenhang mit dem sogenannten Higgs-Mechanismus, wo eine spontane Symmetriebrechung f\"ur die Massen
der $W$- und $Z$-Bosonen verantwortlich gemacht wird. Zudem kann ein solcher Mechanismus auch zur Erkl\"arung des
Unterschiedes zwischen Raum und Zeit herangezogen werden. Offensichtlich k\"onnen unsere Naturgesetze
auf einer vierdimensionalen B\"uhne mit drei r\"aumlichen und einer zeitlichen Dimension formuliert werden,
wenn man von der m\"oglichen Existenz weiterer Dimensionen absieht.
Dass die Zeit sich aber von den drei r\"aumlichen Dimensionen unterscheidet, k\"onnte die Konsequenz eines spontanen
Sym\-metrie\-brechungs\-mecha\-nis\-mus auf einer \"ubergeordneten physikalischen Ebene sein, die sich unserer wissenschaftlichen
Sicht noch entzieht. Wenn Wasser gefriert, m\"ussen sich die im fl\"ussigen Wasser vorwiegend ungeordnet bewegenden Wassermolek\"ule
spontan in r\"aumlich willk\"urlich ausgerichteten Kristallisationsebenen anordnen. In \"ahnlicher Weise k\"onnte die Raumzeit-Struktur von einer
h\"oheren Symmetrieebene ausgehend spontan zur beobachteten Raum- und Zeit-Struktur heruntergebrochen sein
(siehe Dvali et al. 2000 und Referenzen darin).

Man mag sich mit Fug und Recht die Frage stellen, weshalb wir drei Raumdimensionen, doch nur eine Zeitdimension
wahrnehmen k\"onnen. In der Tat w\"are der ``Alltag'' in einer Welt mit zwei Zeitdimensionen recht verwirrend, vor allem
was die Terminplanung betrifft.
Die aus der Existenz nur einer Zeitdimension folgende kausale Struktur unserer Welt ist eine wichtige theoretische
St\"utze, der in der Elementarteilchenphysik eine oft nicht manifest wahrgenommene, aber doch fundamentale Bedeutung
zukommt (Epstein und Glaser 1973). 

\subsubsection*{(Eich-)Formalismen}
Es ist eine empirische Erfahrung, dass unsere Welt einem kontinuierlichen Wandel
unterworfen zu sein scheint. Gehen wir f\"ur den Moment von der Arbeitshypothese aus, dass das Universum ein
zwar komplizierter, aber in gewissem Sinne doch beschreibbarer Mechanismus ist, so stellt sich die Frage,
wie sich die zeitliche Entwicklung der Welt in formaler Weise beschreiben l\"asst.

Die Physiker bedienen sich dazu abstrakter Zustandr\"aume, deren Punkte oder Elemente jeweils dem
Zustand eines physikalischen Systems entsprechen. In der klassischen, also nicht quantisierten Physik arbeitet
man mit Phasenr\"aumen. Der Phasenraum eines idealisiert gedachten Massepunktes beispielsweise, welcher sich in nur
einer Dimension bewegen darf, ist ein zweidimensionaler Raum. Die zwei Koordinaten eines
Punktes in diesem Phasenraum repr\"asentieren dann die Position und den Impuls (entsprechend der Geschwindigkeit)
des Massepunktes und legen so den Zustand des physikalischen Systems ``Massepunkt'' zu einem gegebenen
Zeitpunkt eindeutig fest. Die dynamischen Naturgesetze, oft ausgedr\"uckt durch Differenzialgleichungen, legen
schliesslich fest, wie sich der Systemzustand im Phasenraum zeitlich entwickelt.
 
In der Quantenmechanik arbeitet man mit Hilbertr\"aumen, einer speziellen Klasse von
oft unendlichdimensional gew\"ahlten Vektorr\"aumen. Damit ist der Zustand jedes physikalischen Systems als
Element eines solchen Vektorraumes zu verstehen. Ungl\"ucklicherweise ist die mathematische Situation
in unendlichdimensionalen R\"aumen alles andere als trivial. Dies hat zur Folge, dass es bis dato unm\"oglich war,
eine nachweislich wohldefinierte konsistente Theorie der Elementarteilchen und ihrer Wechselwirkungen zu
formulieren. Dennoch existieren Rechenverfahren, die eine Beschreibung submikroskopischer Prozesse
mit teils erstaunlicher Pr\"azision erlauben.

Ein bei den eben dargestellten Betrachtungen unterschlagenes Problem liegt wiederum im Begriff der
Zeit begr\"undet. Basierend auf Erkenntnissen der allgemeinen Relativit\"atstheorie ist heutzutage sehr wohl bekannt,
dass es eine universelle Zeit, die \"uberall im Universum ``gleichm\"assig tickt'', nicht gibt. Dies bedeutet nicht zwangsl\"aufig,
dass ein Zeitbegriff nicht fassbar gemacht werden k\"onnte, aber doch eine wesentliche Komplikation bei der
konsistenten Formulierung einer Theorie, die die Quanten-Dynamik von Raum und Zeit ber\"ucksichtigen soll.
Die mit dem Zeitbegriff und der Quantentheorie verbundenen theoretischen Fragen werden im Rahmen verschiedenster
Zug\"ange zur Quantengravitationstheorie untersucht, haben aber keine ab\-schlies\-sende Behandlung erfahren.
Wahrscheinlich kommt der Zeitentwicklung des Universums gar keine fundamentale Bedeutung
zu. Die Vergangenheit bestimmt die Zukunft und umgekehrt. Es ist das menschliche Bewusstsein,
welches jeweils nur kleine Ausschnitte aus dem gesamten raum-zeitlich vier- oder noch h\"oher-dimensionalen Leben eines Menschen
fassen kann.

Die Beobachtung, dass der Zustand einer Systems im Rahmen einer formalen Beschreibung 
als Element eines Raumes von alternativen Zust\"anden aufgefasst werden kann, wirft unweigerlich die Frage auf,
ob nicht alle m\"oglichen Zust\"ande in diesem Raum in gewissem Sinne als real aufgefasst werden m\"ussen.
Wenn die Welt eine Welt von vielen denkbaren Welten ist, inwiefern sind nicht alle denkbaren Welten real?
Aus der Sicht moderner Theorien stellt sich weniger die Frage, weshalb etwas ist, sondern vielmehr,
weshalb etwas \emph{nicht ist}. Wir wollen aber den Faden dieser Gedankeng\"ange hier nicht weiterspinnen.
Man bedenke aber, dass die von uns \emph{gerade jetzt} wahrgenommene Welt nicht dieselbe Welt ist wie jene vor einer Sekunde
oder in einer Sekunde. F\"ur den unvoreingenommenen Betrachter hat keine dieser drei Welten einen h\"oheren Realit\"atsanspruch.

Von einer eher technischen Seite her ist zu sagen, dass es sich bei der mathematischen Beschreibung der im SM
auftretenden Wechselwirkungen (oder ``Kr\"afte'') als vorteilhaft erwiesen hat, sogenannte Eichfelder einzuf\"uhren, in
welchen Teilchenzust\"ande eine Rolle spielen, die in der Natur gar nicht beobachtbar sind. In Rahmen
dieser \emph{Eichtheorien} spielen auch sogenannte \emph{Geistteilchen} eine Rolle, deren Aufgabe es ist, die
durch die spezielle Relativit\"atstheorie Einsteins implizierten Raumzeitsymmetrien in den quantenphysikalischen
Formalismus zu integrieren (Faddeev und Popov 1967).
Die Geistteilchen sind unphysikalisch und daher nicht nachweisbar, doch hat
es sich rechnerisch als ein beinahe hoffnungsloses Unterfangen erwiesen, ohne sie vern\"unftig zu arbeiten.

Somit werden die die Wechselwirkungen in der Teilchenphysik beschreibenden Theorien als Eichtheorien
formuliert. Der Bezug zum Wort \emph{Eichung} liegt darin begr\"undet, dass es in einer Theorie letztlich
keine Rolle spielt, wie sie genau formuliert wird, sondern dass die aus ihr errechneten Konsequenzen stimmig
sein m\"ussen. Es ist unwichtig, ob eine Strecke in Metern oder Ellen abgemessen wird, solange die Einheiten korrekt
ineinander umgerechnet werden k\"onnen; ein Kreis kann als eine Menge von Punkten mit konstantem
Abstand von einem Mittelpunkt oder als eine in sich geschlossene Kurve konstanter Kr\"ummung in der
euklidischen Ebene aufgefasst werden. Erstaunlicherweise erm\"oglicht es das Studium der Eich-Freiheiten
bei der Beschreibung einer Elementarteilchentheorie viel \"uber die Theorien selbst zu lernen,
da die Freiheiten, die ein System dem Betrachter bei seiner Beschreibung l\"asst, wiederum R\"uckschl\"usse
auf das System zulassen. Dabei ist es sogar erlaubt, Objekte in die Theorie einzuf\"uhren, die in keiner
unmittelbaren Beziehung zum beschriebenen System stehen.
Eichtheorien sind ein sehr fruchtbares, aber abstraktes Kapitel der modernen Physik, auf
das hier aus Platzgr\"unden nicht weiter eingegangen werden kann (Aste und Scharf 1999).

\subsubsection*{Das Vakuum}
Tats\"achlich spielt in der mathematischen Elementarteilchenphysik ein physikalischer Zustand
eine besondere Rolle, der eine ausserordentlich hohe Symmetrie aufweist.
Man stellt sich vor, dass es einen Zustand der Welt gibt, welcher immer die gleichen Eigenschaften zeigt,
egal ob er verschoben, gedreht oder auf eine gewisse Geschwindigkeit beschleunigt wird.
Dieser Zustand tiefster Energie ist das \emph{Vakuum}.
Dank seinen interessanten Eigenschaften und trotz seines theoretischen Charakters ist es sehr
oft in den Gehirnen theoretischer Physiker zu Gast.

Der leere Raum ist also nicht nichts, sondern die B\"uhne, die durch komplexere Strukturen bev\"olkert
werden kann. Dies impliziert, dass alle Struktur und die Gesetzm\"assigkeiten, denen diese Strukturen
unterliegen, aus diesem Vakuum heraus erzeugt werden k\"onnen. In der Physik wird,
wie in der Einf\"uhrung erw\"ahnt, zu diesem Zwecke
ein mathematisches Konzept verwendet, welches mit sogenannten \emph{Quantenfeldern} operiert.

Auf einer noch fundamentaleren Betrachtungsebene versucht man in modernen Theorien der
Quantengravitation sogar den Aufbau des Raumes selbst zu erkl\"aren. Auch in solchen
Theorien existiert das Konzept des Vakuums, welches dann aber nicht den leeren Raum, sondern
gar die Abwesenheit oder eine minimale Quantit\"at von Raum beschreibt (Ashtekar et al. 2003).

In gewissen Quantenfeldtheorien existieren viele zul\"assige Vakua, von denen eines durch
spontane Symmetriebrechung von der Natur als L\"osung ausgew\"ahlt werden muss.
Dieser Symmetriebruch ist verkn\"upft mit der
Existenz der Masse der Elementarteilchen und wird gemeinhin mit dem bereits erw\"ahnten Ausdruck $Higgs-Mechanismus$
\"Uberschrieben. Kraft eines oder mehrerer ``Higgs-Felder'' wird versucht, eine konsistente Theorie
zu formulieren, in welcher urspr\"unglich masselose und durch diesen Eigenschaftsmangel symmetrischere Objekte durch
Wechselwirkung in massive Objekte transformieren, die man sozusagen auf die Waage legen kann -
bei den Lichtteilchen, den Photonen, welche in solchen Theorien trotz allem masselos bleiben, w\"are dies ein sinnloses Unterfangen.

In popul\"arwissenschaftlichen Darstellungen hat es sich durchgesetzt, das oft unsinnigerweise als
``Gottesteilchen'' bezeichnete Higgs-Teilchen als Urheber
der Masse diverser Teilchen darzustellen. Es existieren aber weitere Mechanismen und
theoretische Modelle, die f\"ur die Massenerzeugung herangezogen werden k\"onnen (Aste et al. 1998).
Zudem kann der Higgs-Formalismus auch so interpretiert werden, dass die Existenz massiver Teilchen
die Einf\"uhrung von Higgs-Teilchen erfordert, um die Konsistenz der zugrundeliegenden Quantenfeldtheorien
sicherzustellen (Aste et al. 1999). Schliesslich ist zu bemerken, dass ein Elektron als genauso g\"ottlich oder
teuflisch wie ein Higgs-Teilchen bezeichnet werden kann.

\subsubsection*{Lokalisierbarkeit}
Es ist eines des Resultate der Quantenphysik, dass die Frage nach dem pr\"azisen Aufenthaltsort
eines Teilchens im Allgemeinen nicht scharf definiert ist. Diese Unbestimmtheit, welche durch die
ber\"uhmte \emph{Heisenbergsche Unsch\"arferelation} quantitativ erfasst werden kann, hat mit der
eigentlichen geometrischen Ausdehnung eines Teilchens nicht unmittelbar zu tun. Vielmehr unterscheiden sich
die sinnvollen Messgr\"ossen, die mit einem physikalischen System verkn\"upft sind, in der klassischen Physik und
der Quantenmechanik. Die Frage nach einer exakten aktuellen Position eines Teilchens ist zu vergleichen
mit der Frage, welches Gewicht die Kreiszahl $\pi$ besitzt. Die Lokalisierungseigenschaften eines Teilchens
k\"onnen innerhalb gewisser Grenzen durch \emph{Wellenfunktionen} beschrieben
werden, aus welchen sich Wahrscheinlichkeiten berechnen lassen, ein Teilchen in einem gewissen
Raumzeitbereich anzutreffen.

Tats\"achlich ist die Frage nach der Definition des Teilchenbegriffs heute noch mit mathematischen
Spitzfindigkeiten und Fragen behaftet. Ein geladenes Teilchen wie das Elektron besitzt bekanntlicherweise
ein es umgebendes elektrisches Feld. Wird das Elektron beschleunigt, so reagiert auf diese \"Anderung
auch das Feld, allerdings mit zeitlicher Verz\"ogerung. Die Frage, inwiefern das Elektron von seinem
elektrischen Feld zu unterscheiden ist, f\"uhrt zu erheblichen theoretischen Komplikationen, die zwar
f\"ur praktische Rechnungen umgangen werden k\"onnen, aber letztlich nicht gel\"ost sind (Schroer 2008).

Der Gesamtzustand eines physikalischen Systems wird, wie bereits erw\"ahnt, durch einen
Vektor in einem passend gew\"ahlten Hilbertraum dargestellt. Dieser Vektor ist zugleich synonym zur oben erw\"ahnten
Wellenfunktion. Die Eigenschaften des physikalischen Systems,
welche sich aus dem Zustand ableiten lassen, folgen aus einem mathematischen Apparat, in welchem
wiederum sogenannte \emph{Quantenfelder} eine Rolle spielen. Diese Quantenfelder, in welchen die physikalischen
Naturgesetze in gewisser Weise kodiert sind, gingen historisch durch einen Abstraktionsprozess aus den
Wellenfunktionen hervor.

W\"ahrend in der \emph{Ph\"anomenologie} als Sparte der theoretischen Teilchenphysik
eher pragmatisch und unter Vernachl\"assigung ganz rigoroser Begr\"undungen versucht wird, durch teils grossen Rechenaufwand
Voraussagen und Interpretationen f\"ur Experimente zu erm\"oglichen, wird im Rahmen der \emph{axiomatischen
Quantenfeldtheorie} versucht, die Theorie auf eine echte mathematisch konsistente Basis zu stellen
(Haag 1992). Bei diesem Unterfangen treten praktische Anwendungen eher in den Hintergrund.

\subsubsection*{Supersymmetrie}
Nach der sehr wahrscheinlichen Entdeckung des im SM postulierten Higgs-Teilchens im Jahre 2012 am CERN
(CERN 2012) in Genf hoffen viele Physiker auf die Entdeckung weiterer Teilchen, welche im Rahmen
des SM noch nicht beschrieben werden. Das entdeckte Teilchen ist etwa so schwer wie ein Bariumatom,
welches in seinem Kern 56 Protonen und noch mehr Neutronen enth\"alt, und gilt doch in gewissem Sinne als
elementar.

Es gibt gute Gr\"unde anzunehmen, dass das SM in einigen Teilen nicht konsistent ist und daher erweitert werden
muss. Viele bisher rein hypothetische Erweiterungen des SM beinhalten eine neuartige Symmetrie, welche
als \emph{Supersymmetrie} ($SUSY$) bezeichnet wird. Diese Symmetrie stellt eine Beziehung zwischen Fermionen und
Bosonen her, \"ahnlich der $CPT$-Symmetrie, welche eine Beziehung zwischen Teilchen und ihren Antiteilchen vermittelt.
Aufgrund einer spontanen Symmetriebrechung ist die $SUSY$ in unserer Welt aber nicht exakt realisiert.
Auf jedes bisher bekannte Boson kommt rechnerisch ein Fermion als supersymmetrisches Partnerteilchen,
ein so genanntes Bosino. Die ``Superpartner'' der Bosonen werden durch die Endung
\emph{-ino} im Namen gekennzeichnet, so heisst beispielsweise das dem (hypothetischen)
Gluon entsprechende Fermion Gluino. Entsprechende hypothetische Superpartner existieren auch zu den bereits bekannten Fermionen.
Den Quarks werden dadurch Squarks zugeordnet, Leptonen erhalten Sleptonen als Partner. Die tats\"achliche Zuordnung ist
aber kompliziert, da Quantenfelder miteinander ``vermischt'' werden k\"onnen. Zudem sagen die Modelle mehrere Teilchen
voraus, die als Higgs-Teilchen aufgefasst werden k\"onnen. Keines der schon bekannten Bosonen ist ein $SUSY$-Partner
eines bereits bekannten Fermions.

Bisher wurde aber keines der postulierten supersymmetrischen Partnerteilchen experimentell nachgewiesen.
Diese m\"ussen Eigenschaften wie beispielsweise eine so hohe Masse haben, dass sie unter normalen Bedingungen nicht entstehen.
Man hofft noch, dass Teilchenbeschleuniger wie der \emph{Large Hadron Collider} am CERN zumindest einige dieser
Teilchen direkt oder indirekt nachweisen k\"onnen. Mit dem leichtesten dieser supersymmetrischen Teilchen hofft man zudem, einen
Kandidaten f\"ur die vermutete dunkle Materie des Universums zu finden, welche noch unbekannter Natur ist und doch einen gr\"osseren
Anteil als die uns bekannten Formen der Materie im Universum beitr\"agt.

\subsection*{Literaturverzeichnis}
\begin{description}
\item Ashtekar,~A., Bojowald,~M., Lewandowski,~J. (2003):
Mathematical structure of loop quantum cosmology. Advances in Theoretical and Mathematical
Physics 7, 233-268. 
\item Aste,~A. (2013), Scharf,~G. (1999):
Non-abelian gauge theories as a consequence of perturbative quantum gauge invariance.
International Journal of Modern Physics A14, 3421-3434. 
\item Aste,~A., Scharf,~G., D\"utsch,~M. (1999):
Perturbative gauge invariance: electroweak theory II. 
Annalen der Physik 8, 389-404.
\item Aste,~A., Scharf,~G., von Arx,~C. (2010):
Regularization in quantum field theory from the causal point of view.
Progress in Particle and Nuclear Physics 64, 61-119.
\item Aste,~A., Scharf,~G., Walther,~U. (1998):
Power counting degree versus singular order in the Schwinger model. Nuovo Cimento  A111, 323-327.
\item Canetti,~L., Drewes,~M., Shaposhnikov,~M. (2012): Matter and Antimatter in the Universe.
New Journal of Physics 14, 095012.
\item CMS Collaboration (2012): Observation of a new boson at a mass of 125 GeV with the CMS experiment at the LHC.
Physics Letters B716, 30-61. 
\item D\"utsch,~M., Gracia-Bondia,~J. (2012): On the assertion that PCT violation implies Lorentz
non-invariance. Physics Letters B711, 428-433.
\item  Dvali,~G., Gabadadze,~G., Porrati,~M. (2000): 
4-D gravity on a brane in 5-D Minkowski space.
Physics Letters B485, 208-214.
\item Epstein,~H., Glaser,~V. (1973):
The role of locality in perturbation theory.
Annales de l'institut Henri Poincar\'e (A) Physique th\'eorique 19, 211-295.
\item Faddeev,~V., Popov,~L. (1967): Feynman diagrams for the Yang-Mills field. Physics Letters B25, 29-30.
\item Haag,~R. (1992): Local Quantum Physics: Fields, Particles, Algebras. Springer-Verlag, Berlin, Heidelberg, New York.
\item Lie,~S., Engel,~F., (1888): Theorie der Transformationsgruppen. Verlag B.G.~Teubner, Leipzig.
\item L\"uders, G. (1957): Proof of the TCP theorem. Annals of Physics (New York) 2, 1-15.
\item Pauli,~W. (1925): \"Uber den Zusammenhang des Abschlusses der Elektronengruppen im Atom mit
der Komplexstruktur der Spektren. Zeitschrift f\"ur Physik 31, 765-783.
\item Schroer,~B. (2008): A note on infraparticles and unparticles. arXiv:0804.3563.
\item Wu,~C., Ambler,~E., Hayward,~R., Hoppes,~D., Hudson,~R. (1957): Experimental Test of
Parity
Conservation in Beta Decay. Physical Review 105, 1413-1415.
\end{description}
}
\end{multicols}

\end{document}